\documentclass[onecolumn,showpacs,amsmath,amssymb,prd,nofootinbib,floatfix,preprintnumbers]{revtex4-2}

\usepackage[dvipdfmx]{graphicx}
\usepackage{dcolumn}
\usepackage{bm}
\usepackage{hyperref}
\usepackage{comment}
\usepackage{color}
\usepackage[rgb]{xcolor}
\usepackage{amsmath}
\usepackage[normalem]{ulem}
\newcommand{\bx}{{\mathbf{x}}}

\newcommand{\bk}{{\mathbf{k}}}
\newcommand{\bd}{{\mathbf{d}}}

\newcommand{\br}{{\mathbf{r}}}

\newcommand{\hx}{\hat{x}}

\newcommand{\hk}{\hat{k}}

\newcommand{\he}{\hat{e}}

\newcommand{\hr}{\hat{r}}
\newcommand{\hz}{\hat{z}}

\newcommand{\ha}{\hat{a}}
\newcommand{\hb}{\hat{b}}
\newcommand{\hu}{\hat{u}}
\newcommand{\hw}{\hat{w}}
\newcommand{\hbx}{\hat{\mathbf{x}}}

\newcommand{\hbk}{\hat{\mathbf{k}}}
\newcommand{\hbd}{\hat{\mathbf{d}}}
\newcommand{\hbe}{\hat{\mathbf{e}}}

\newcommand{\hbn}{\hat{\mathbf{n}}}
\newcommand{\hbr}{\hat{\mathbf{r}}}

\newcommand{\hba}{\hat{\mathbf{a}}}
\newcommand{\hbb}{\hat{\mathbf{b}}}
\newcommand{\hbu}{\hat{\mathbf{u}}}
\newcommand{\hbw}{\hat{\mathbf{w}}}

\definecolor{olivegreen}{rgb}{0,0.6,0}
\definecolor{mycyan}{rgb}{0.13, 0.62, 0.8}
\definecolor{deepcarminepink}{rgb}{0.94, 0.19, 0.22}

\newcommand{\mnras}{Mon.~Not.~Roy.~Astron.~Soc.}
\newcommand{\physrep}{Phys.~Rep.}
\newcommand{\jcap}{JCAP}
\newcommand{\pasj}{Publ.~Astron.~Soc.~Japan}

\newcommand{\aap}{Astronomy~\&~Astrophysics}

\newcommand{\ssr}{Space~Science~Rev.}

\newcommand{\avrg}[1]{\left\langle #1 \right\rangle}

\newcommand{\rvred}[1]{\textcolor{black}{#1}}

\begin{document}

\preprint{IPMU~22-0001}

\title{
Analysis method for 3D power spectrum of projected tensor field 
with fast estimator and window convolution modelling: 
an application to intrinsic alignments
}

\author{Toshiki Kurita$^{1,2}$}\email{toshiki.kurita@ipmu.jp}
\author{Masahiro Takada$^{1}$}
\affiliation{%
$^1$Kavli Institute for the Physics and Mathematics of the Universe (WPI),\\
 The University of Tokyo Institutes for Advanced Study (UTIAS),\\
 The University of Tokyo, Chiba 277-8583, Japan
}%
\affiliation{%
$^2$Department of Physics, Graduate School of Science,\\
 The University of Tokyo, 7-3-1 Hongo, Bunkyo-ku, Tokyo 113-0033, Japan
}%

\date{\today}

\begin{abstract}
Rank-2 tensor fields of large-scale structure, e.g. a tensor field inferred from {\it shapes} of galaxies, open up a window to directly access 2-scalar, 2-vector and 2-tensor modes, where the scalar fields can be measured independently from the standard density field that is traced by distribution of galaxies. 
Here we develop an estimator of the multipole moments of coordinate-independent power spectra for the three-dimensional tensor field, taking into account the projection of the tensor field onto plane perpendicular to the line-of-sight direction.
To do this, we find that a convenient representation of the power spectrum multipoles can be obtained by the use of the associated Legendre polynomials in the form which allows for the fast Fourier transform estimations under the local plane-parallel (LPP) approximation. 
The formulation also allows us to obtain the Hankel transforms to connect the two-point statistics in Fourier and configuration space, which are needed to derive theoretical templates of the power spectrum including convolution of a survey window. 
To validate our estimators, we use the simulation data of the projected tidal field assuming a survey window that mimics the BOSS-like survey footprint. 
We show that  the LPP estimators fairly well recover the multipole moments that are inferred from the global plane-parallel approximation. 
We find that the survey window causes a more significant change in the multipole moments of projected tensor power spectrum at $k\lesssim 0.1\,h{\rm Mpc}^{-1}$ from the input power spectrum, than in the density power spectrum. 
Nevertheless, \rvred{our method to compute the theory template}
including the survey window effects successfully 
\rvred{reproduces}
the window-convolved multipole moments measured from the simulations. 
The analysis method presented here paves the way for a cosmological analysis using three-dimensional tensor-type tracers of large-scale structure for current and future surveys.
\end{abstract}
\maketitle

\section{\label{sec:introduction}Introduction}

There are ongoing and upcoming wide-area cosmology surveys such as the Subaru Hyper Suprime-Cam (HSC) survey \citep{2018PASJ...70S...4A}, the Subaru Prime Focus Spectrograph (PFS) \citep{2014PASJ...66R...1T}, the Dark Energy Spectrograph Instrument (DESI)\footnote{\url{https://www.desi.lbl.gov}}, ESA Euclid, Rubin Observatory's Legacy Survey of Space and Time (LSST)\footnote{\url{https://lsst.slac.stanford.edu}} and NASA Roman Space Telescope. 
These surveys will enable to address fundamental questions of the universe such as the nature of dark matter and dark energy and the physics involved in generation of primordial fluctuations that are the seeds of cosmic structures today, with unprecedented precision. 
A standard method used to study large-scale structure (LSS) is based on statistics of ``density'' field of LSS tracers. 
For example, the three-dimensional spatial distribution of galaxies, inferred by observed angular positions and photometric or spectroscopic redshifts of individual galaxies, has been used to constrain cosmological parameters of the standard $\Lambda$CDM model \cite[e.g.][]{2005ApJ...633..560E,2021arXiv211006969K}, properties of the primordial perturbations \citep[e.g.,][]{2008PhRvD..77l3514D,2022arXiv220107238C}, and gravity theories beyond general relativity \citep[e.g.][]{2010Natur.464..256R}. 

Generally speaking, we can also explore vector- and tensor-type components from LSS observables, which carry complementary or even independent cosmological information from that from the scalar-type observables. 
For example, the redshift-space distortion (RSD) effect \citep{1987MNRAS.227....1K} and the kinetic Sunyaev-Zel'dovich effect \citep{1972CoASP...4..173S} arise from peculiar velocities of galaxies or galaxy clusters, which could in general contain the information on vector-type components, in addition to the information of scalar gravitational potential. 
Also promisingly, {\it shapes} of galaxies, which are usually quantified by the ellipticities, can be used to extract rank-2 tensor information in large-scale structure. 
Physical correlations between shapes of different galaxies, the so-called intrinsic alignments (IA), have been studied as one of important effects in large-scale structure physics \citep{Croft&Metzler2000:IA_dawn,Catelan+2001:IA_dawn,Crittenden+2002:IA_dawn,Jing&Suto2002:IA_dawn}  \citep[also see][for for reviews]{Joachimi+2015:IA_review,Kirk+2015:IA_review,Kiessling+2015:IA_review,Troxel&Ishak2015:IA_review}.

The leading theory for the IA effect is the linear alignment model (LA) \citep{Hirata&Seljak2004:IA_LA}, which predicts that galaxy shapes originate from the primordial tidal field in large-scale structure. 
This model has also been extended to include the effect from nonlinear structure formation \citep{Bridle&King2007:IA_NLA,Blazek+2017:IA_TATT,Schmitz+2018:IA_TATT,Vlah+2020:IA_EFT,Schneider&Bridle2010:IA_HM,Fortuna+2020:IA_HM}. 
Here the nonlinear evolution generally induces vector- and tensor-type modes in the observed IA effect on small scales; e.g., the nonlinear effect induces $B$-mode in the IA signal on small scales, even if the IA effect starts from a pure $E$-mode (scalar tidal field) in the linear regime.  
There are also numerical studies to investigate IA; 
using $N$-body simulations for dark matter halo shapes \citep{Jing2002:IA_nbody,Xia+2017:IA_nbody,Piras+2018:IA_nbody,Osato+2018:IA_nbody,Okumura+2019:IA_nbody,Okumura+2020:IA_nbody,Kurita+2020:IA_nbody,Stucker+2021:IA_SU,Akitsu+2021:IA_SU};
using hydrodynamical simulations for simulated galaxy shapes \citep{Tenneti+2015a:IA_hydro,Tenneti+2015b:IA_hydro,Chisari+2015:IA_hydro,Velliscig+2015:IA_hydro,Chisari+2016:IA_hydro,Hilbert+2017:IA_hydro,Chisari+2017:IA_hydro,Shi+2020:IA_hydro,Samuroff+2021:IA_hydro,Shi+2021:IA_hydro}.
In an observation side, many works \citep{Mandelbaum+2006:IA_measurement,Okumura+2009:IA_measurement,Singh+2015:IA_measurement,Samuroff+2018:IA_measurement,Yao+2020:IA_measurement,Fortuna+2021:IA_measurement,Tonegawa&Okumura2021:IA_measurement} have reported significant detections of IA correlation functions from actual dataset such as luminous red galaxies in the SDSS dataset. 

Although the IA effect has been studied mainly as a contamination to weak lensing, recently several works have proposed that IA can be used as a new cosmological probe. 
For example, Refs.~\citep{Chisari&Dvorkin2013:IA_PNG,Schmidt+2015:IA_PNG,Kogai+2018:IA_PNG} proposed that scale-dependent shape bias in the large-scale IA effect can be used to explore the spin-2 (quadrupolar) anisotropic local-type primordial non-Gaussianity (PNG), independently from the effect of the isotropic local-type PNG on the galaxy bias at large scales \citep{2008PhRvD..77l3514D}.
This PNG effect on the IA power spectrum was confirmed using halo shapes in $N$-body simulations where the anisotropic PNG initial conditions are implemented \citep{Akitsu+2021:IA_PNG}.
These are interesting directions to explore because the PNG signal gives a smoking-gun signal of \rvred{the} physics in the early universe such as inflation \citep{2003JHEP...05..013M}. 
Also interestingly, characteristic tensor-type (therefore $B$-mode) signatures can be induced by the primordial gravitational wave \citep{Schmidt&Jeong2012:LSSwithGW_2,Schmidt+2014:LSSwithGW_3} due to the so-called fossil effect on large-scale structure.
Furthermore, even within the standard $\Lambda$CDM model, Refs.~\citep{Okumura&Taruya2020:IA_improvement,Taruya&Okumura2020:IA_improvement} discussed that a measurement of the IA effect can improve cosmological parameter estimation when combined with the galaxy clustering correlations. 

Motivated by the above background, in this paper we develop a method to measure the three-dimensional power spectrum of rank-2 tensor tracers of LSS from a realistic cosmology survey. 
To have a practical method that can be applied to actual data, we consider the tensor field that can be obtained by projecting the rank-2 tensor field onto local 2D plane perpendicular to the line-of-sight (LOS) direction to each LSS tracer (e.g. galaxy). 
This is indeed the case for galaxy shapes that are measured from the projected light distribution (i.e. surface brightness distribution). 
Since the projected tensor field varies with the LOS direction, we will first define coordinate-independent power spectra of the projected tensor field. 
To keep generality of our formulation, we will express the projected tensor power spectrum in terms of the underlying power spectra of scalar-, vector- and tensor modes and derive
how the derived power spectra are related to the coordinated-independent power spectra or the $E$- and $B$-mode power spectra, where the $E/B$-mode decomposition is useful for the projected rank-2 field, e.g. because the scalar perturbation induces only the $E$ mode in the linear regime and the $B$ mode gives a smoking-gun signature of vector- or tensor-type perturbation.

Furthermore, extending the so-called Yamamoto estimator \citep{Yamamoto+2006:estimator} or the local plane-parallel (LPP) approximation \citep{Scoccimarro2015:estimator,Bianchi+2015:estimator,Hand+2017:estimator} for redshift-space power spectrum of galaxies, we will develop an estimator of the projected tensor power spectra. 
To do this, we will use the {\it associated} Legendre polynomials, instead of the standard Legendre polynomials, to define multipole moments of the projected tensor power spectra. 
This allows us to derive a convenient representation of the multipole moments that allows for fast Fourier transform estimation of the power spectrum moments from an actual wide-area survey. 
We will also use the representation to derive the Hankel transforms to relate the multipole moments in Fourier and configuration space. 
We will use the Hankel transforms to derive equations for computing theoretical template of the projected tensor power spectra including the effect of survey window convolution, in analogy with the form of the standard galaxy power spectrum \citep{Wilson+2017}.
Then we will use simulations of the tidal field assuming the BOSS-like survey footprint to demonstrate validation of our method for the LPP estimator of the projected tensor power spectrum and the accuracy of theoretical templates including the survey window convolution. 
With this study, we will be ready to apply our method to actual data such as the SDSS data, e.g. for constraining the anisotropic PNG signal in the rank-2 tensor tracers of LSS.

This paper is organized as follows.
In Section~\ref{sec:observables}, we first define helicity-based decomposition of rank-2 tensor field and then derive formula for the coordinate-independent power spectra of the projected tensor field. 
In Section~\ref{sec:methodology}, we develop a methodology of the power spectrum analysis for the projected tensor field including FFT-based LPP estimators and survey window convolution.
In Section~\ref{sec:validation}, we use the simulated tidal field to validate our method assuming the BOSS-like survey geometry. 
Finally we give some discussion along an application of our method to real data in Section~\ref{sec:discussion} and then give conclusion in Section~\ref{sec:conclusion}.

Throughout this paper, we use the following abbreviations:
\begin{align*}
    \int_\bx \equiv \int \mathrm{d}\bx,~
    \int_\bk \equiv \int \frac{\mathrm{d}\bk}{(2\pi)^3},~
    \int_{\hbk} \equiv \int \frac{\mathrm{d}\Omega_{\hbk}}{4\pi},
\end{align*}
and we use quantities with hat symbol $(\hat{\hspace{0.5em}})$ to denote their unit vectors. 
In addition we define the Fourier and inverse Fourier transforms as
\begin{align*}
    f(\bk) \equiv \int_\bx f(\bx) e^{-i\bk\cdot\bx},~
    f(\bx) \equiv \int_\bk f(\bk) e^{i\bk\cdot\bx}.
\end{align*}
and denote the Hankel and inverse Hankel transforms in terms of the $\ell$-th order spherical Bessel function as
\begin{align*}
    \mathcal{H}_\ell \left[ g(r) \right](k) &\equiv 4\pi(-i)^\ell \int r^2 \mathrm{d}r j_{\ell}(kr) g(r), \\
    \mathcal{H}^{-1}_\ell \left[ g(k) \right](r) &\equiv i^{\ell} \int \frac{k^2 \mathrm{d}k}{2\pi^2} j_{\ell}(kr) g(k).
\end{align*}

\section{Tensor Observables of Large-scale Structure}
\label{sec:observables}

\subsection{Tensor Field}
\label{subsec:tensor_field}

In this paper we consider a 3D tensor field estimated from large-scale structure observables, whose components are denoted as $s_{ij}(\bx)$ satisfying $s_{ij}=s_{ji}$. 
The Fourier transform is given by $s_{ij}(\bk) \equiv \int_{\bx}~ s_{ij}(\bx)~ e^{-i\bk \cdot \bx}$. 
An example is the tensor field that can be inferred from shapes of galaxies or dark matter halos.
The galaxy or halo shape is characterized by the second moments of their light or mass density profile, which form the tensor field sampled at galaxy/halo positions.
The rank-2 tensor field $s_{ij}$ carries 6 components at each position, which correspond to scalar, vector and tensor modes, respectively, as discussed in the following.
The tensor field is, without loss of generality, can be decomposed into 6 orthogonal polarization states, which we label $p=\{0, z, x, y, +, \times\}$ following the notations in Ref.~\citep{Donghui&Marc2012:fossil}:
$s_{ij}(\bk) \equiv \sum_p \bar{s}^{(p)}(\bk) \bar{\varepsilon}^{(p)}_{ij}(\hbk)$ where $\hbk$ is the unit vector of $\bk$ and $\bar{\varepsilon}^{(p)}_{ij}(\hbk)$ is the polarization basis. 
The 6 polarization bases are; 
$\bar{\varepsilon}^{(0)}_{ij} = \sqrt{1/3} \delta^K_{ij}$ is the scalar mode of the trace component where $\delta^K_{ij}$ is the Kronecker delta function;    
$\bar{\varepsilon}^{(z)}_{ij} = \sqrt{3/2}( \hk_i\hk_j - \delta^K_{ij}/3)$ is the longitudinal scalar mode
($\bar{\varepsilon}^{(z)}_{ii}=0$); 
$\bar{\varepsilon}^{(x,y)}_{ij} = \sqrt{1/2}(\hk_i \hw^{(x,y)}_j + \hk_j \hw^{(x,y)}_i)$
are the vector modes, where $\hbw^{(x,y)}(\hbk)$ are two orthonormal and transverse vectors satisfying 
$\hbw^{(p)} \cdot \hbw^{(p')} = \delta^K_{pp'}$ and $\hbk\cdot\hbw^{(p)}=0$;
$\bar{\varepsilon}^{(+,\times)}_{ij} = \sqrt{1/2}(\hw^{(x)}_i\hw^{(x)}_j-\hw^{(y)}_i\hw^{(y)}_j, \hw^{(x)}_i\hw^{(y)}_j+\hw^{(x)}_j\hw^{(y)}_i)$ are two traceless and \rvred{transverse} tensor modes which satisfy
$\bar{\varepsilon}^{(+,\times)}_{ii}=0$ and $\bar{\varepsilon}^{(+,\times)}_{ij} \hk_j = 0$.
Note that the results we will show below are independent of the specific choice of $\hbw^{(x,y)}$.
We define the normalization factor of each basis so that $\bar{\varepsilon}^{(p)}_{ij} \bar{\varepsilon}^{(p')}_{ij} = \delta^K_{pp'}$.
For a galaxy shape case the trace component $\bar{s}^{(0)}(\bk)$ characterizes a size of galaxy. 
The trace component would be generally an independent observable from the traceless scalar component \citep[see e.g.][]{Schmidt+2009:size,Joachimi+2015:FP,Singh+2020:FP}. 
However the analysis method of the trace component is the same as that for the density fluctuation field, which is well studied for a galaxy clustering analysis.
Hence from now on, we denote the trace component simply as $\delta(\bx)$ and focus on other 5 traceless components. 

As more convenient expressions, we alternatively expand the tensor field by their transformation properties under a rotation around $\hbk$ as
\begin{align}
    s_{ij}(\bk) = \sum_{m=-2}^{2} s^{(m)}(\bk)~ \varepsilon^{(m)}_{ij}(\hbk),
    \label{eq:helicity_decomposition_of_sij}
\end{align}
where the new basis $\left\{ \varepsilon^{(m)}_{ij} \right\}$ is the so-called helicity basis related to the polarization basis as
\begin{align}
    \varepsilon^{(0)}_{ij} = \bar{\varepsilon}^{(z)}_{ij},~
    \varepsilon^{(\pm 1)}_{ij} = \frac{1}{\sqrt{2}} \left( \bar{\varepsilon}^{(x)}_{ij} \pm i \bar{\varepsilon}^{(y)}_{ij} \right),~ 
    \varepsilon^{(\pm 2)}_{ij} = \frac{1}{\sqrt{2}} \left( \bar{\varepsilon}^{(+)}_{ij} \pm i \bar{\varepsilon}^{(\times)}_{ij} \right).
\end{align}
Note that the helicity tensors transform as $\varepsilon^{(m)}_{ij} \to e^{im\psi} \varepsilon^{(m)}_{ij}$ under a rotation around $\hbk$ by angle $\psi$.
Requiring statistical homogeneity and isotropy and parity invariance, we can write the auto-power spectra of the tensor field as a sum over power spectra of each helicity field \citep{Vlah+2020:IA_EFT}:
\begin{align}
    \langle s_{ij}(\bk)~ s_{kl}(\bk') \rangle \equiv (2\pi)^3 \delta^3_D(\bk+\bk') 
    \left\{ \Lambda^{(0)}_{ij,kl}(\hbk)~ P_{ss}^{(0)}(\bk) + \sum_{\lambda=1}^2 \Lambda^{(\lambda)}_{ij,kl}(\hbk)~ \frac{P_{ss}^{(\lambda)}(\bk)}{2} \right\}, 
    \label{eq:def_of_full_tensor_auto_spectrum}
\end{align}
or equivalently,
\begin{align}
    \langle s_{ij}(\bk)~ s_{ij}(\bk') \rangle \equiv (2\pi)^3 \delta^3_D(\bk+\bk') 
    \sum_{\lambda=0}^2 P_{ss}^{(\lambda)}(\bk),
\end{align}
where $\delta^3_D(\bk)$ is Dirac delta function and $\Lambda^{(\lambda)}_{ij,kl} \equiv \sum_{m=\pm \lambda}  \varepsilon^{(m)}_{ij}\varepsilon^{(m)*}_{kl}$ are the rotationally invariant tensors in the plane normal to $\hbk$ with respect to the helicity $\lambda \equiv |m|$, and the explicit forms are: 
\begin{align}
    \Lambda^{(0)}_{ij,kl}(\hbk) &= \frac{3}{2} \left( \hk_i\hk_j - \frac{1}{3}\delta^K_{ij} \right) \left( \hk_k\hk_l - \frac{1}{3}\delta^K_{kl} \right), \label{eq:Lambda0_def}\\
    \Lambda^{(1)}_{ij,kl}(\hbk) &= \frac{1}{2} \left( \mathcal{P}_{ik}(\hbk)\hk_j\hk_l + \mathcal{P}_{il}(\hbk)\hk_j\hk_k + \mathcal{P}_{jk}(\hbk)\hk_i\hk_l + \mathcal{P}_{jl}(\hbk)\hk_i\hk_k \right), \label{eq:Lambda1_def}\\
    \Lambda^{(2)}_{ij,kl}(\hbk) &= \frac{1}{2} \left( \mathcal{P}_{ik}(\hbk)\mathcal{P}_{jl}(\hbk) + \mathcal{P}_{il}(\hbk)\mathcal{P}_{jk}(\hbk) - \mathcal{P}_{ij}(\hbk)\mathcal{P}_{kl}(\hbk) \right),
    \label{eq:Lambda2_def}
\end{align}
where $\mathcal{P}_{ij}(\hba) \equiv \delta^K_{ij} - \ha_i \ha_j$ is the projection tensor with respect to $\hba$.

In addition to the auto-power spectra, we define the cross-power spectra with the density field as
\begin{align}
    \langle s_{ij}(\bk)~ \delta(\bk') \rangle \equiv (2\pi)^3 \delta^3_D(\bk+\bk') \varepsilon^{(0)}_{ij}(\hbk)~ P_{s\delta}^{(0)}(\bk). 
    \label{eq:def_of_full_cross_power}
\end{align}

\subsection{Projected Tensor Field}
\label{subsec:projected_tensor_field}

It is usually difficult to estimate the 3D tensor field from large-scale structure observables. 
What is easier to measure instead is a {\it projected} tensor field, where the projection is onto plane perpendicular to the LOS direction at each position. 
For a galaxy shape case, we can estimate the projected shape of a galaxy from its observed surface brightness distribution.
Correlations of the projected shape field, i.e. IA correlations \citep{Catelan+2001:IA_dawn,Hirata&Seljak2004:IA_LA}, have been observed from actual data \citep[e.g.,][]{Singh+2015:IA_measurement,Singh&Mandelbaum2016:IA_measurement}. 
In this paper we do not discuss another source of correlations of observed galaxy shapes, which arise from cosmological weak lensing effects due to the foreground large-scale structure (the so-called cosmic shear), so please do not confuse IA with cosmic shear. 
We hereafter denote the projected tensor field as $\gamma_{ij}(\bx)$ satisfying $\hx_i \gamma_{ij}(\bx) = 0$, where we set an observer's position to the coordinate origin and $\hbx$ becomes the unit vector along the line-of-sight direction to $\bx$ (in this case $\hat{x}^i$ is equivalent to the angular position of $\bx$ on the celestial sphere of an observer in our choice of the coordinate). 
The projected tensor field can be defined by
\begin{align}
    \gamma_{ij}(\bx) \equiv 
    \Lambda^{(2)}_{ij,kl} (\hbx)
    s_{kl}(\bx).
    \label{eq:2Dshear_from_3Dshape}
\end{align}
Note that the last term of $\Lambda^{(2)}_{ij,kl}$ in Eq.~(\ref{eq:Lambda2_def}) ensures that $\gamma_{ij}$ is traceless; $\gamma_{ii}=0$. 
Hence $\gamma_{ij}$ carries two modes. 
The coordinated-independent two modes, often used in the literature, are $E$- and $B$-modes, as discussed below.

For later convenience, we introduce the complex representation of the projected tensor field using the helicity basis in real space: 
\begin{align}
    _{\pm 2}\gamma(\bx) \equiv \gamma_1(\bx) \pm i \gamma_2(\bx) \equiv e^{(\pm2)}_{ij}(\hbx) 
    \gamma_{ij}(\bx) = e^{(\pm 2)}_{ij}(\hbx) s_{ij}(\bx),
    \label{eq:gamma_def}
\end{align}
where $e^{(\pm 2)}_{ij}$ is defined in terms of the orthonormal basis; $\{ \hbx, \hbe_\theta, \hbe_\phi \}$ as
\begin{align}
    e^{(\pm 2)}_{ij}(\hbx) \equiv \he^{(\pm 1)}_{i}(\hbx) \he^{(\pm 1)}_j(\hbx), 
    \label{eq:def_of_eij_+2}
\end{align}
with
\begin{align}
    \hbe^{(\pm 1)}(\hbx) \equiv \frac{1}{\sqrt{2}}\left(\hbe_\theta \pm i \hbe_\phi \right),
    \label{eq:def_polarization_vector}
\end{align}
and $\hbe_\theta$ and $\hbe_\phi$ are the unit vectors in the 2D plane perpendicular to the line-of-sight direction $\hbx$, e.g. the RA and Dec directions in the celestial coordinate.
We have used the relation $e^{(\pm2)}_{ij} \Lambda^{(2)}_{ij,kl}  = e^{(\pm2)}_{kl}$ in the last equality of Eq.~(\ref{eq:gamma_def}). 
The helicity basis $e^{(\pm 2)}_{ij}$ satisfies identities, 
$e^{(\pm 2)}_{ii}=e^{(\pm 2)}_{ij}e^{(\pm 2)}_{ij}=0$ and $e^{(\pm 2)}_{ij}e^{(\mp 2)}_{ij}=1$.
The projected tensor components $\gamma_1$ and $\gamma_2$ correspond to the distortions along the directions of coordinate axes ($\hbe_\theta$ or $\hbe_\phi$) and the directions rotated by $45^{\circ}$ from coordinate axes, respectively \citep[also see Ref.][for the definition]{Schmidt&Jeong2012:LSSwithGW_2}.
Hereafter we will use only the ``$+2$'' component because the results in the following sections are identical to those in the case of using the ``$-2$'' component;
we will omit the label ``$+2$'' for notational simplicity, e.g. $\gamma(\bx) \equiv {}_{+2}\gamma(\bx)$, $e_{ij}(\hbx) \equiv e^{(+2)}_{ij}(\hbx)$ from here on.

\subsection{Power Spectrum of Projected Tensor Field}
\label{sec:gpp}

In this section we briefly review derivation of power spectra 
of the projected tensor field assuming the distant observer approximation or equivalently the global plane-parallel (GPP) approximation.

As we described above (Eq.~\ref{eq:2Dshear_from_3Dshape}), the projected tensor field, $\gamma_{ij}(\bx)$, is obtained from projection of $s_{ij}(\bx)$ onto plane perpendicular to the LOS direction ($\hbx$). 
Hence the projected components of tensor field varies with the LOS direction, and its Fourier transform is generally given by a convolution form as $\gamma_{ij}(\bk) = \int_{\bk'}\Lambda^{(2)}_{ij,kl}(\bk-\bk')\, s_{kl}(\bk')$.
Assuming the GPP approximation, the tensor field shares the same, global LOS direction, which we denote as a constant unit vector $\hbn$ in the following.
In this case the projection tensor becomes a constant tensor, and the helicity basis in real space, $e_{ij}$, and the orthonormal vectors, $\hbe_{\theta}$ and $\hbe_{\phi}$, become independent of $\hbx$.
The projected tensor field reduces to $\gamma_{ij}(\bx; \hbn) \simeq \Lambda^{(2)}_{ij,kl}(\hbn) s_{kl}(\bx)$ and also its Fourier transform is simply expressed by a multiplication form as $\gamma_{ij}(\bk; \hbn) \simeq \Lambda^{(2)}_{ij,kl}(\hbn) s_{kl}(\bk)$. 

Since the complex projected tensor field in Fourier space is given as $\gamma(\bk; \hbn) \equiv e_{ij}(\hbn) s_{ij}(\bk)$, we can construct the so-called $E$/$B$-mode fields as
\begin{align}
    \Gamma(\bk; \hbn) &\equiv E(\bk; \hbn) + iB(\bk; \hbn) \equiv \gamma(\bk; \hbn)e^{-2i\phi_{\hbk,\hbn}},
\end{align}
where the phase factor is given as
\begin{align}
    e^{2i\phi_{\hbk,\hbn}} \equiv \frac{2e_{ij}(\hbn)\hk_i\hk_j}{\mathcal{P}_{ij}(\hbn)\hk_i\hk_j}, 
    \label{eq:def_of_phase_factor}
\end{align}
which determines a rotation of the shear components on the plane perpendicular to the LOS direction by angle between $\hbe_\theta(\hbn)$ and the projected wavevector; $k_{\perp i} \equiv \mathcal{P}_{ij}(\hbn) k_j$.
Using these modes, we can define the {\it coordinate-independent} power spectra from the observed fields $\delta$ and $\Gamma$ as
\begin{align}
    \avrg{\Gamma(\bk) \delta^*(\bk')} &\equiv (2\pi)^3 \delta^3_D(\bk-\bk') P_{\gamma \delta}(\bk) \equiv (2\pi)^3 \delta^3_D(\bk-\bk') P_{E \delta}(\bk), \label{eq:def_of_prj_cross_power}\\
    \avrg{\Gamma(\bk) \Gamma^*(\bk')} &\equiv (2\pi)^3 \delta^3_D(\bk-\bk') P_{+}(\bk) \equiv (2\pi)^3 \delta^3_D(\bk-\bk') [P_{EE}(\bk)+P_{BB}(\bk)], \label{eq:def_of_prj_auto_power_+}\\
    \avrg{\Gamma(\bk) \Gamma(\bk')} 
    &\equiv (2\pi)^3 \delta^3_D(\bk-\bk') P_{-}(\bk) \equiv (2\pi)^3 \delta^3_D(\bk-\bk') [P_{EE}(\bk)-P_{BB}(\bk)], \label{eq:def_of_prj_auto_power_-}
\end{align}
where we have introduced the $E$- and $B$-mode power spectra, $P_{EE}$ and $P_{BB}$, and we have assumed that the imaginary part corresponding to parity-odd power spectrum vanishes due to parity invariance: $\avrg{B\delta}=\avrg{EB}=0$. 
The power spectra generally depend on the norm of Fourier modes $k\equiv |\bk|$ and the angle, $\mu_k \equiv \hbk \cdot \hbn$; $P(\bk) = P(k,\mu_k)$.

We can relate the projected tensor power spectra to the full 3D power spectra of the tensor field (see Appendix~\ref{sec:relation_full_and_projected_spectrum} for the derivation):
\begin{align}
    P_{\gamma \delta}(\bk) &= \sqrt{\frac{3}{8}} (1-\mu_k^2) P_{s\delta}^{(0)}(\bk), \label{eq:3D2Drelation_Pcross}\\
    P_{+}(\bk) &= \frac{3}{8} (1-\mu_k^2)^2 P_{ss}^{(0)}(\bk) + \frac{1}{8} (1-\mu_k^2)\{ (1-\mu_k)^2 + (1+\mu_k)^2\} P_{ss}^{(1)}(\bk) + \frac{1}{32} \{ (1-\mu_k)^4 + (1+\mu_k)^4 \} P_{ss}^{(2)}(\bk), \label{eq:3D2Drelation_P+}\\
    P_{-}(\bk) &= (1-\mu_k^2)^2 \left[ \frac{3}{8} P_{ss}^{(0)}(\bk) - \frac{1}{4} P_{ss}^{(1)}(\bk) + \frac{1}{16} P_{ss}^{(2)}(\bk) \right]. \label{eq:3D2Drelation_P-}
\end{align}
The spectra, $P^{(0)}_{s\delta}$ and $P^{(\lambda)}_{ss}$ ($\lambda=0,1,2$), generally depend on $|\bk|$ and $\mu_k$, e.g. due to the RSD effect \citep{1987MNRAS.227....1K}, the primordial fossil effects \citep{Donghui&Marc2012:fossil}, and also the super-survey tidal effect \citep{2017PhRvD..95h3522A,2018PhRvD..97f3527A}. 
Hence we use vector notation $\bk$ in the argument of power spectra to keep generality of our discussion.
We should stress that the $\mu_k$-dependent prefactors in front of the underlying tensor power spectra on the r.h.s., such as $(1-\mu_k^2)$, are purely from geometrical effects due to the projection from $s_{ij}$ to $\gamma_{ij}$.
In particular, the factorized form of the cross- and ``minus''-power spectra become important when we construct FFT-based estimators of the power spectra of the projected tensor field and derive the associated Hankel transforms that give the 1D integral relations to the correlation functions.

The $E$- and $B$-mode auto-power spectra of the projected tensor field are explicitly given as
\begin{align}
    P_{EE} &= \frac{1}{2} \left( P_{+} + P_{-} \right) = \frac{3}{8} (1-\mu_k^2)^2 P_{ss}^{(0)} + \frac{1}{4}\mu_k^2(1-\mu_k^2) P_{ss}^{(1)} + \frac{1}{16} (1+\mu_k^2)^2 P_{ss}^{(2)}, \nonumber\\
    P_{BB} &= \frac{1}{2} \left( P_{+} - P_{-} \right) = \frac{1}{4} (1-\mu_k^2) P_{ss}^{(1)} + \frac{1}{2} \mu_k^2P_{ss}^{(2)}. 
    \label{eq:relation_pm_EB}
\end{align}
Thus the $E$-mode power spectrum arises from the scalar (helicity-0), vector (helicity-1) and tensor (helicity-2) modes, while the $B$-mode spectrum is from the vector and tensor modes, as in the CMB polarization power spectra \citep{1997PhRvD..55.1830Z,1997PhRvD..55.7368K,1997PhRvD..56..596H}, the cosmic shear spectra \citep{2002ApJ...568...20C,2002A&A...389..729S,2003MNRAS.344..857T}, and the IA spectra \citep{2001ApJ...559..552C,Kurita+2020:IA_nbody}.
Note that Ref.~\citep{Vlah+2021:IA_EFT} employed the Limber approximation to derive the $E/B$-mode angular power spectra of the IA shear, where only the Fourier modes with $\mu_k=0$ are considered, and arrived at conclusion that the $B$-mode angular power spectrum is only from the vector mode.

\subsection{Hankel Transforms}
\label{subsec:connection_with_correlation_function}

In this section we derive the Hankel transforms to relate the power spectra of the projected tensor field to the two-point correlation functions. 
We will later use the Hankel transform expressions to derive the formula of power spectrum including the effect of survey window convolution. 

Let us begin with defining coordinate-independent correlation functions of the projected tensor field.
For this purpose we introduce the projected tensor field at the position $\bx$, 
defined with respect to line connecting $\bx$ and $\bx'$, denoted as $\gamma_r(\bx;\bx')$:
\begin{align}
    \gamma_r(\bx; \bx') \equiv \gamma_+(\bx; \bx') + i \gamma_\times(\bx; \bx') \equiv \gamma(\bx)e^{-2i\phi_{\hbr,\hbn}},
\end{align}
with
\begin{align}
    e^{2i\phi_{\hbr,\hbn}} \equiv \frac{e_{ij}(\hbn)\hr_i\hr_j}{\mathcal{P}_{ij}(\hbn)\hr_i\hr_j},
    \label{eq:phase_r}
\end{align}
where $\br \equiv \bx - \bx'$, the vector connecting $\bx$ and $\bx'$.
$\phi_{\hbr, \hbn}$ in Eq.~(\ref{eq:phase_r}) is the phase factor rotating the tensor components on the plane perpendicular to the LOS direction and thus $\gamma_+ (\gamma_\times)$ is the tangential (cross) component with respect to the projected relative vector; $\br_{\perp i} \equiv \mathcal{P}_{ij}(\hbn)\br_j$\footnote{In the case of $(\hbn, \hbe_\theta, \hbe_\phi) = (\hbx_3,\hbx_1,\hbx_2)$ for instance, we reproduce the standard matrix representation as
\begin{align}
    \begin{pmatrix}
        \gamma_+ \\
        \gamma_\times \\
    \end{pmatrix} 
    =
    \begin{pmatrix}
        \cos{2\phi} & \sin{2\phi} \\
        -\sin{2\phi} & \cos{2\phi} \\
    \end{pmatrix} 
    \begin{pmatrix}
        \gamma_1 \\
        \gamma_2 \\
    \end{pmatrix},
\end{align}
where $\phi \equiv \phi_{\hbr, \hbx_3}$ is the angle between $\br_\perp$ and $\hbx_1$}.

Using the above fields, we can now define the coordinate-independent correlation functions as
\begin{align}
    \xi_{\gamma \delta}(\br) &\equiv \avrg{\gamma_r(\bx; \bx') \delta(\bx')} = \avrg{\gamma(\bx) \delta(\bx')}e^{-2i\phi_{\hbr,\hbn}}, \label{eq:def_of_prj_cross_corr}\\
    \xi_+(\br) &\equiv \avrg{\gamma_r(\bx; \bx')\gamma_r^*(\bx'; \bx)} = \avrg{\gamma(\bx)\gamma^*(\bx')}, 
    \label{eq:def_of_prj_auto_corr_+}\\
    \xi_-(\br) &\equiv \avrg{\gamma_r(\bx; \bx')\gamma_r(\bx'; \bx)} = \avrg{\gamma(\bx)\gamma(\bx')}e^{-4i\phi_{\hbr,\hbn}}. \label{eq:def_of_prj_auto_corr_-}
\end{align}
These correlation functions are free of choices of the coordinates $\hbe_\theta$ and $\hbe_\phi$; e.g. a rotation of the $\hbe_\theta$ and $\hbe_\phi$ coordinates does not change these correlation functions. 
The correlation functions generally depend on the distance between two points, $r \equiv |\br|$, and angle, $\mu_r \equiv \hbr \cdot \hbn$; $\xi(\br) = \xi(r,\mu_r)$, e.g. due to the RSD effect. 

Using the definitions of the power spectra in Eqs.~(\ref{eq:def_of_prj_cross_power})--(\ref{eq:def_of_prj_auto_power_-}) and those of the correlation functions in Eqs.~(\ref{eq:def_of_prj_cross_corr})--(\ref{eq:def_of_prj_auto_corr_+}), we can find relations to connect these as\footnote{The expressions of Eqs.~(\ref{eq:relation_cross_gpp})--(\ref{eq:relation_auto_gpp_-}) are in analogy with the 2D (angular) statistics.
If we replace the 3D vectors, $(\br,\bk)$ with the 2D vectors, $(\boldsymbol{\theta},\boldsymbol{\ell})$ in Eqs.~(\ref{eq:relation_cross_gpp})--(\ref{eq:relation_auto_gpp_-}), the expressions just correspond to the well-known relation between the angular power spectrum and the angular 2D correlation function under the global plane-parallel approximation: e.g. 
\begin{align*}
    C_{\gamma \delta}(\ell) 
    = \int_{\boldsymbol{\theta}} \xi^{2D}_{\gamma \delta}(\boldsymbol{\theta}) e^{2i(\phi_{\boldsymbol{\theta}} - \phi_{\boldsymbol{\ell}})} e^{-i\boldsymbol{\ell} \cdot \boldsymbol{\theta}} 
    = 2\pi \int \theta \mathrm{d}\theta \xi^{2D}_{\gamma \delta}(\theta) J_{2}(\ell\theta),
\end{align*}
where $J_m$ is the Bessel function of order $m$.}
\begin{align}
    P_{\gamma \delta}(\bk) &= \int_\br \xi_{\gamma \delta}(\br) e^{2i(\phi_{\hbr,\hbn} - \phi_{\hbk,\hbn})} e^{-i\bk \cdot \br}, \label{eq:relation_cross_gpp}\\
    P_{+}(\bk) &= \int_\br \xi_+(\br) e^{-i\bk \cdot \br}, \label{eq:relation_auto_gpp_+}\\ 
    P_{-}(\bk) &= \int_\br \xi_-(\br) e^{4i(\phi_{\hbr,\hbn} - \phi_{\hbk,\hbn})} e^{-i\bk \cdot \br}.
    \label{eq:relation_auto_gpp_-}
\end{align}

We consider multipole moments of the power spectrum, which is defined by the angle average with respect to 
$\hbk$, and derive the Hankel transform.
The easiest case is the Hankel transform for the ``plus'' auto-power spectrum, $P_+(\bk)$ because it is similar to that for the redshift-space power spectrum of the density field: 
\begin{align}
    P^{(\ell)}_{+}(k) &\equiv (2\ell+1) \int_{\hbk}~P_+(\bk){\cal L}_{\ell}(\mu_k)\nonumber\\
    &=4\pi (-i)^\ell \int r^2 \mathrm{d}r~ \xi^{(\ell)}_{+}(r) j_\ell(kr) 
    \left( = \mathcal{H}_{\ell}\left[\xi^{(\ell)}_+(r)\right] (k) \right),
    \label{eq:Hankel_trs_+}
\end{align}
where ${\cal L}_\ell(x)$ is the $\ell$-th order Legendre polynomial, and we have defined multipole moments of the power spectrum and correlation function as $P_{+}(\bk) = \sum_{\ell} P^{(\ell)}_{+}(k) \mathcal{L}_{\ell} (\mu_k)$ and $\xi_{+}(\br) = \sum_{\ell} \xi^{(\ell)}_{+}(r) \mathcal{L}_{\ell} (\mu_r)$. 

On the other hand, in the case of the cross-power spectrum $P_{\gamma\delta}$ and the ``minus'' auto-power spectrum $P_-$, there are additional phase factors and thus the angle integration is not trivial unlike the density auto or ``plus'' auto power spectra.
To overcome this complexity, we use the {\it associated} Legendre polynomials, instead of ${\cal L}_{\ell}$, which are defined as 
\begin{align}
    \mathcal{L}^{m}_{L}(\mu) \equiv (-1)^m (1-\mu^2)^{m/2} \frac{{\rm d}^m}{{\rm d}\mu^m} \mathcal{L}_{L}(\mu).
    \label{eq:def_of_asso}
\end{align}
The associated Legendre polynomials satisfy the following orthogonal relation and the integration identity (see Appendix~\ref{sec:derivations} for the proof): 
\begin{align}
 \int_{-1}^{1}\frac{\mathrm{d}\mu}{2}~{\cal L}_{L}^m(\mu){\cal L}_{L'}^m(\mu)&=\frac{(L+m)!}{(2L+1)(L-m)!}\delta^K_{LL'} \label{eq:ortho_asso_legendre}\\
 \int_{\hbk} e^{-im\phi_{\hbk,\hbn}} \mathcal{L}^{m}_{L}(\hbk \cdot \hbn) \mathcal{L}_{\ell'}(\hbk \cdot \hbr) &= \frac{1}{2L+1}\delta^K_{L\ell'} \mathcal{L}^{m}_{L}(\hbr \cdot \hbn) e^{-im\phi_{\hbr,\hbn}}, \label{eq:formulae_associatedLegendre}
\end{align}
Throughout this paper we use the capital letter ``$L$'' to denote the order of associated Legendre polynomials (${\cal L}^m_{L})$, while we use the lower letter ``$\ell$'' to denote the order of Legendre polynomials (${\cal L}_\ell$).
Let us first consider the cross power spectrum (Eq.~\ref{eq:relation_cross_gpp}). 
Introducing the multipole moments of the cross spectrum and the cross correlation, expanded in terms of the associated Legendre polynomials with $m=2$, as
\begin{align}
    P^{(L)}_{\gamma \delta}(k)&\equiv (2L +1 )\frac{(L-2)!}{(L+2)!}\int_{\hbk}P_{\gamma\delta}(\bk){\cal L}^{m=2}_{L}(\hbk\cdot\hbn),  \label{eq:multipole_cross_def} \\
    \xi_{\gamma\delta}(\br)&\equiv \sum_{L=2} \xi^{(L)}_{\gamma\delta}(r){\cal L}^{m=2}_{L}(\hbr\cdot\hbn),
\end{align}
we can find the following the Hankel transform (see Appendix~\ref{sec:hankel_transform}): 
\begin{align}
    P^{(L)}_{\gamma\delta}(k) = 4\pi (-i)^L \int r^2 \mathrm{d}r~ \xi^{(L)}_{\gamma\delta}(r) j_L(kr) 
    \left( = \mathcal{H}_{L}\left[\xi^{(L)}_{\gamma \delta}(r)\right] (k) \right), 
    \label{eq:Hankel_trs_cross}
\end{align}
for integer $L$ with $L\ge 2$, and we introduced the factor $(L-2)!/(L+2)!$ in Eq.~(\ref{eq:multipole_cross_def}) accounting of the orthogonality relation of the associated Legendre polynomials (Eq.~\ref{eq:ortho_asso_legendre}). 
Also note $P_{\gamma\delta}(\bk)=\sum_{L=2}P^{(L)}_{\gamma\delta}(k){\cal L}^{m=2}_{L}(\hbk\cdot\hbn)$.
Thus the associated Legendre polynomials with order $m=2$ can nicely deal with the phase factor in Eq.~(\ref{eq:relation_cross_gpp})\footnote{We can expand the cross statistics in terms of the usual Legendre polynomials instead as in the case of the galaxy clustering; 
$P_{\gamma\delta}(\bk)=\sum_{\ell=0}P^{(\ell)}_{\gamma\delta}(k){\cal L}_{\ell}(\hbk\cdot\hbn)$ and $\xi_{\gamma\delta}(\br) = \sum_{\ell=0} \xi^{(\ell)}_{\gamma\delta}(r){\cal L}_{\ell}(\hbr\cdot\hbn)$, however, the moments of the same order $\left(P^{(\ell)}_{\gamma\delta},\xi^{(\ell)}_{\gamma\delta} \right)$ are ${\it not}$ directly associated with each other by the Hankel and inverse Hankel transformations.}.
The cross power spectrum of projected tensor field is particularly important, e.g. for IA measurements, so this is one of the main results of this paper.

Similarly, using the associated Legendre polynomials with order $m=4$, we can find the Hankel transform for the minus auto-power spectrum of shear as
\begin{align}
    P^{(L)}_{-}(k) &\equiv  (2L+1)\frac{(L-4)!}{(L+4)!}\int_{\hbk}~P_{-}(\bk){\cal L}^{m=4}_{L}(\hbk\cdot\hbn)\nonumber\\
    &= 4\pi (-i)^{L} \int r^2 \mathrm{d}r~ \xi^{(L)}_{-}(r) j_{L}(kr)
    \left( = \mathcal{H}_{L}\left[\xi^{(L)}_{-}(r)\right] (k) \right), 
    \label{eq:Hankel_trs_-}
\end{align}
for integer $L$ with $L\ge 4$, and note $P_-(\bk)=\sum_{L=4}~P^{(L)}_{-}(k){\cal L}^{m=4}_{L}(\hbk\cdot\hbn)$.

Before proceeding we would like to notice the following things. 
One might think that Eqs.~(\ref{eq:Hankel_trs_cross}) and (\ref{eq:Hankel_trs_-}) are not a complete expansion of the underlying power spectra, $P_{\gamma\delta}$ and $P_-$, at first glace, because a set of the associated Legendre polynomials $\left\{ \mathcal{L}^{m}_{L}(\mu) \right\}$ with fixed $m(\neq0)$ is not a complete \rvred{basis}. 
An arbitrary function $f(\mu)$ cannot be generally expanded by a series of $\mathcal{L}^{m}_{L}(\mu)$; e.g., $f(\mu)=1$ cannot be expanded by $\{{\cal L}^{m=2}_L\}$, because the lowest order ${\cal L}^{m=2}_L$ is ${\cal L}^{m=2}_{L=2}(\mu)=3(1-\mu^2)$. 
However, this is not true. 
As can be found from Eqs.~(\ref{eq:3D2Drelation_Pcross}) and (\ref{eq:3D2Drelation_P-}), the power spectra $P_{\gamma\delta}$ and $P_-$ are given by the underlying 3D power spectra of the tensor fields, $P^{(0)}_{s\delta}$ and $P_{ss}^{(\lambda)}$, with geometrical prefactors $(1-\mu_k^2)$ and $(1-\mu_k^2)^2$, respectively: 
$P_{\gamma\delta}\propto(1-\mu_k^2)P^{(0)}_{\delta s}$ and $P_-\propto(1-\mu_k^2)^2\left[P_{ss}^{(0)}+\cdots\right]$. 
Recalling the definitions $P_{\gamma\delta}(\bk)=\sum_{L=2}P^{(L)}_{\gamma\delta}(k){\cal L}^{m=2}_{L}(\hbk\cdot\hbn)$ and 
$P_{-}(\bk)=\sum_{L=4}P^{(L)}_{-}(k){\cal L}^{m=4}_{L}(\hbk\cdot\hbn)$, we can find that a set of $\left\{{\cal L}^{m=2}_L(\mu_k)/(1-\mu_k^2)\right\}$ and $\left\{{\cal L}^{m=4}_L(\mu_k)/(1-\mu_k^2)^2\right\}$ includes a complete set of $\mu$-polynomials, $\left\{\mu_k^0,\mu_k^1,\mu_k^2,\cdots \right\}$. 
Hence a set of the multipole moments, $P_{\gamma\delta}^{(L)}(k)$ or $P_{-}^{(L)}(k)$, carries the full information on the underlying power spectra, $P_{s\delta}^{(0)}(\bk)$ and $P_{ss}^{(\lambda)}(\bk)$. 

Using the Hankel transform expressions Eqs.~(\ref{eq:Hankel_trs_cross}) and (\ref{eq:Hankel_trs_-}), we will below develop a method that allows an efficient FFTlog based computation of the power spectrum including the survey window effect. 
Note that Refs.~\citep{Okumura&Taruya2020:IA_improvement,Okumura+2020:IA_nbody} derived the Hankel transform relations between the 3D power spectrum and the 3D correlation function for the IA shear field assuming the linear IA model and linear RSD effect. 
Compared to this, we have not employed any model for the projected tensor field or the underlying power spectra. 
Therefore the above Hankel transform formulae are valid for any projected tensor field.

\section{Methodology of Analysis}
\label{sec:methodology}

\subsection{Estimators}
\label{subsec:estimator}

In a wide solid-angle survey, the projection to define the projected tensor field varies with the LOS direction ($\hbx$) as we discussed around Eq.~(\ref{eq:gamma_def}).
In this section we construct estimators for multipole moments of 3D power spectrum of such {\it LOS-dependent} projected tensor field, with the form allowing fast Fourier transforms (FFTs). 

\subsubsection{Galaxy clustering}
\label{subsubsec:review_clustering}

Before going to the power spectra of projected tensor field, we briefly review a derivation for FFT-based estimators for multipole moments of redshift-space galaxy power spectrum, following \citet{Yamamoto+2006:estimator} \citep[also see Refs.][]{Scoccimarro2015:estimator,Bianchi+2015:estimator}. 

Since the RSD effect arises from the LOS component of the peculiar velocity of each galaxy in pair, the redshift-space power spectrum breaks statistical translation invariance.
Even in such case, we can define the multipole moments around an observer by the spatial average of the local power spectrum considering the LOS directions toward the local regions \citep[see][for more detailed discussion]{Scoccimarro2015:estimator}: 
\begin{align}
    \rvred{\hat{P}}^{(\ell)}(k) 
    &\equiv (2\ell+1) \int_{\hbk,\bx,\bx'} \delta(\bx) \delta(\bx') e^{-i\bk \cdot (\bx - \bx')} \mathcal{L}_\ell (\hbk \cdot \hbd),
    \label{eq:lpp_dens}
\end{align}
where $\bx$ and $\bx'$ are positions of paired two galaxies used in the two-point correlation function estimate, and $\bd = \bd (\bx,\bx')$ is the direction referring to the pair $(\bx,\bx')$\rvred{, e.g. the midpoint vector of the two galaxies; $\bd \equiv (\bx+\bx')/2$}. 
Note that, when all the galaxies in a survey region is sufficiently distant to an observer or we can assume the so-called distant observer approximation, we can take one global $\hbn$ for the LOS direction to all galaxy pairs and Eq.~(\ref{eq:lpp_dens}) reduces to the GPP estimator: 
\begin{align}
    \rvred{\hat{P}}^{(\ell)}(k) \simeq (2\ell+1) \int_{\hbk,\bx,\bx'} \delta(\bx) \delta(\bx') e^{-i\bk \cdot (\bx - \bx')} \mathcal{L}_\ell (\hbk \cdot \hbn) = (2\ell+1) \int_{\hbk} |\delta(\bk)|^2 \mathcal{L}_\ell (\hbk \cdot \hbn).
    \label{eq:gpp_dens}
\end{align}
This recovers the statistical translation invariance. 

The direct estimation using Eq.~(\ref{eq:lpp_dens}) is computationally expensive for $\ell>0$ because we need to calculate the double sum with respect to $\bx$ and $\bx'$ due to $\bd(\bx,\bx')$.
To avoid this problem, we approximate $\hbd$ as the direction toward one galaxy in each pair, i.e. $\hbd \simeq \hbx$ in Eq.~(\ref{eq:lpp_dens}) which is so-called the endpoint approximation \citep{Bianchi+2015:estimator, Scoccimarro2015:estimator, Hand+2017:estimator, Oliver&Zachary2021:estimator}. 
The estimator (Eq.~\ref{eq:lpp_dens}) can be reduced to 
\begin{align}
    \rvred{\hat{P}}^{(\ell)}(k)
    &\simeq (2\ell+1) \int_{\hbk,\bx,\bx'} \delta(\bx) \delta(\bx') e^{-i\bk \cdot (\bx - \bx')} \mathcal{L}_\ell (\hbk \cdot \hbx) \nonumber \\
    &= (2\ell+1) \int_{\hbk} \left[ \int_{\bx} \delta(\bx) e^{-i\bk \cdot \bx} \mathcal{L}_\ell (\hbk \cdot \hbx) \right] \left[ \int_{\bx'} \delta(\bx') e^{i\bk \cdot \bx'} \right] \nonumber\\
    &\equiv (2\ell+1) \int_{\hbk} \delta^{(\ell)}(\bk)~ \delta(-\bk), \label{eq:lpp_dens_app}
\end{align}
where
\begin{align}
    \delta^{(\ell)}(\bk) \equiv \int_{\bx} \delta(\bx) e^{-i\bk \cdot \bx} \mathcal{L}_\ell (\hbk \cdot \hbx).
    \label{eq:delta_ell}
\end{align}
Note that $\delta^{(0)}(\bk) = \delta(\bk)$.
This is often called Yamamoto estimator or the local plane-parallel (LPP) estimator for galaxy power spectrum.
One advantage of this approximation is that we can use the FFT algorithm to compute $\delta^{(\ell)}(\bk)$ decomposing the Legendre polynomial in Eq.~(\ref{eq:delta_ell}) into the sum of the products of Cartesian components \citep{Bianchi+2015:estimator, Scoccimarro2015:estimator}, e.g., $\mathcal{L}_2 (\hbk \cdot \hbx) = \frac{3}{2}\hk_i\hk_j\hx_i\hx_j - \frac{1}{2}$, or the spherical harmonics \citep{Hand+2017:estimator}:
\begin{align}
    \mathcal{L}_\ell (\hbk \cdot \hbx) = \frac{4\pi}{2\ell+1} \sum_{m=\ell}^\ell Y^m_\ell(\hbk) Y^{m*}_\ell(\hbx),
    \label{eq:spherical_hamornics_hand}
\end{align}
thereby allowing an efficient, fast computation of the multipole moments.
The LPP estimator (Eq.~\ref{eq:lpp_dens_app}) has been found to be fairly accurate on scales of interest, compared to the estimator (Eq.~\ref{eq:lpp_dens}) \citep{Scoccimarro2015:estimator,Bianchi+2015:estimator}.

\subsubsection{Projected tensor field}
\label{subsubsec:lpp_for_prj_tensor}

In this section we construct estimators for multipole moments of power spectrum of the projected tensor field.
In doing this, we need to take into account the LOS-direction dependence of the projection operator $\mathcal{P}_{ij}(\hbx)$ in each galaxy pair.

Let us first consider the cross power spectrum, $P_{\gamma\delta}$. 
From Eqs.~(\ref{eq:def_of_prj_cross_corr}), (\ref{eq:relation_cross_gpp}) and (\ref{eq:multipole_cross_def}), we can formally define the estimator for multipole moments of $P_{\gamma\delta}$, without the GPP approximation, in analogy to Eq.~(\ref{eq:lpp_dens}): 
\begin{align}
    \rvred{\hat{P}}^{(L)}_{\gamma \delta}(k) 
    \equiv (2L+1) \frac{(L-2)!}{(L+2)!} \int_{\hbk, \bx,\bx'} \gamma(\bx) \delta(\bx') e^{-2i\phi_{\hbk,\hbd}} e^{-i\bk \cdot (\bx - \bx')} \mathcal{L}^{m=2}_{L}(\hbk \cdot \hbd),
    \label{eq:lpp_estimator_cross}
\end{align}
where we recast the definition of the phase factor for later use: 
\begin{align}
    e^{2i\phi_{\hbk,\hbd}} &\equiv \frac{2e_{ij}(\hbd)\hk_i\hk_j}{\mathcal{P}_{ij}(\hbd)\hk_i\hk_j} = \frac{2e_{ij}(\hbd)\hk_i\hk_j}{1-(\hbk \cdot \hbd)^2}.
\end{align}
Note that we replaced the vector argument of ${\cal L}^{m=2}_L$, $\hbn$, in Eq.~(\ref{eq:multipole_cross_def}) with $\hbd$ in Eq.~(\ref{eq:lpp_estimator_cross}) to explicitly include the LOS dependence, instead of the global (constant) LOS direction. 

Using the endpoint approximation, $\hbd \simeq \hbx$, similarly to Eq.~(\ref{eq:lpp_dens_app}) and writing the phase factor explicitly, Eq.~(\ref{eq:lpp_estimator_cross}) reduces 
\begin{align}
    \rvred{\hat{P}}^{(L)}_{\gamma \delta}(k) 
    &\simeq (2L+1) \frac{(L-2)!}{(L+2)!} \int_{\hbk} \left[ \int_{\bx} \gamma(\bx) 2e^*_{ij}(\hbx) e^{-i\bk \cdot \bx} \frac{\mathcal{L}^{m=2}_L (\hbk \cdot \hbx)}{1-(\hbk \cdot \hbx)^2} \right] \hk_i\hk_j \left[ \int_{\bx'} \delta(\bx') e^{i\bk \cdot \bx} \right] \nonumber \\
    &\equiv (2L+1) \frac{(L-2)!}{(L+2)!} \int_{\hbk} \Xi^{(L)}_{ij}(\bk) \hk_i\hk_j \delta(-\bk), \label{eq:lpp_estimator_cross_Xiij}
\end{align}
where
\begin{align}
    \Xi^{(L)}_{ij}(\bk) \equiv \int_{\bx} \gamma(\bx) 2e^*_{ij}(\hbx) e^{-i\bk \cdot \bx} \frac{\mathcal{L}^{m=2}_L (\hbk \cdot \hbx)}{1-(\hbk \cdot \hbx)^2}.
    \label{eq:def_of_Xiij}
\end{align}
In the first equality on the r.h.s. of Eq.~(\ref{eq:lpp_estimator_cross_Xiij}), we were able to rewrite the estimator by a product form of the Fourier transforms as explicitly denoted by the square brackets, as in Eq.~(\ref{eq:lpp_dens_app}). 
In particular, the use of the associated Legendre polynomials of order $m=2$ leads to a cancellation of the factor in the denominator of one Fourier transform, $1-(\hbk\cdot\hbx)^2$, because ${\cal L}^{m=2}_L(\hbk\cdot\hbx)$ has the overall same factor $1-(\hbk\cdot\hbx)^2$ at all order: 
consequently, ${\cal L}^{m=2}_L(\hbk\cdot\hbx)/[1-(\hbk\cdot\hbx)^2]$ becomes polynomials of $(\hbk\cdot\hbx)$ as in the standard Legendre polynomials, or similarly to Eq.~(\ref{eq:lpp_dens_app}). 
Thus the product form allows us to use FFTs. 
For example, since $\Xi^{(L)}_{ij}=\Xi^{(L)}_{ji}$ (therefore 6 components), we need to compute a total of 
$6\times(1+5+9)=90$ FFTs for the multipole moments with $L=2,4,6$.

Similarly, assuming the LPP endpoint approximation, we can define the estimators for the multipole moments of the auto-power spectra, $P_+$ and $P_-$, from Eqs.~(\ref{eq:relation_auto_gpp_+}) and (\ref{eq:relation_auto_gpp_-}) as explicitly derived in Appendix~\ref{sec:calc_auto_est}:
\begin{align}
    \rvred{\hat{P}}^{(\ell)}_+(k) &\equiv (2\ell+1) \int_{\hbk} \gamma^{(\ell)}(\bk)~ \gamma^{*}(-\bk),
    \label{eq:lpp_estimator_auto_+} \\
    \rvred{\hat{P}}^{(L)}_-(k)
    &\equiv (2L+1) \frac{(L-4)!}{(L+4)!} \int_{\hbk} \Xi^{(L)}_{ijkl}(\bk) \hk_i\hk_j \hk_k\hk_l~ \gamma(-\bk).
    \label{eq:lpp_estimator_auto_-}
\end{align}
where $\gamma^{*}(-\bk)\equiv \int_{\bx}\gamma^*(\bx)e^{i\bk\cdot\bx}$ and we have defined the auxiliary fields as
\begin{align}
    \gamma^{(\ell)}(\bk) &\equiv \int_{\bx} \gamma(\bx) e^{-i\bk \cdot \bx} \mathcal{L}_{\ell} (\hbk \cdot \hbx), \\
    \Xi^{(L)}_{ijkl}(\bk) &\equiv \int_{\bx} \gamma(\bx) 4e^*_{ij}(\hbx) e^*_{kl}(\hbx) e^{-i\bk \cdot \bx} \frac{\mathcal{L}^{m=4}_L (\hbk \cdot \hbx)}{[1-(\hbk\cdot\hbx)^2 ]^2}.
    \label{eq:def_of_Xiijkl}
\end{align}
Note that as shown in Eq.~(\ref{eq:relation_auto_gpp_+}), the relation between the ``plus'' power spectrum, $P_+$, and the correlation function, $\xi_+$, obeys the same rule as in the density case.
Therefore the multipole components of it can be defined in terms of the usual Legendre polynomials and thus the resulting estimator, Eq.~(\ref{eq:lpp_estimator_auto_+}), reduces to the similar form of the clustering estimator, Eq.~(\ref{eq:lpp_dens_app}).
On the other hand, in the case of the ``minus'' power spectrum, $P_-$, the additional phase factor $e^{-4i\phi}$ exists in the transformation relation (Eq.~\ref{eq:relation_auto_gpp_-}) as in the case of the cross-power spectrum.
Hence we define the multipole coefficients with respect to the associated Legendre polynomials of order 4, $\mathcal{L}^{m=4}_L$.
Again the integrand in Eq.~(\ref{eq:def_of_Xiijkl}) is the product form of $\hbx$ and $\hbk$ thanks to the cancellation of $[1-(\hbk\cdot\hbx)^2 ]^2$.
Since $\Xi^{(L)}_{ijkl}$ has symmetric properties, it requires a total of $15\times(1+5+9)=225$ FFT computations for the multipoles with $L=4,6,8$.

\subsection{Window Convolutions}
\label{subsec:window_convolution}

For an actual observation the power spectrum measurement is affected by survey window effects.
In this section, we derive formula for the window-convolved power spectrum of the projected tensor field, in analogy with the window convolution for the power spectrum of galaxy density field \citep{Wilson+2017,Beutler+2017,Beutler&McDonald2021:wide_angle}.
The detail derivation is shown in Appendix~\ref{sec:window_convolution} and here we show the key equations.

We can measure the underlying cosmology fields through 
a survey window function: 
\begin{align*}
    \tilde{\delta}(\bx) &= W_\delta(\bx)\delta(\bx), \\
    \tilde{\gamma}(\bx) &= W_\gamma(\bx)\gamma(\bx),
\end{align*}
where $W_\delta, W_\gamma$ are the window functions for the density field and projected tensor field, respectively.
We use the tilde symbol to denote the observed fields.
The window functions $W_\delta$ and $W_\gamma$ are generally different, but in the following we assume $W_\delta(\bx) = W_\gamma(\bx) \equiv W(\bx)$ for simplicity.
For convenience, we also define the the auto-correlation of the window function:
\begin{align}
    Q(\br) \equiv \int_{\bx} W(\bx) W(\bx+\br),
    \label{eq:windowQ}
\end{align}
and its multipole moments under the LPP approximation are defined as
\begin{align}
    Q_{\ell}(r) \equiv (2\ell+1) \int \frac{\mathrm{d}\Omega_{\hbr}}{4\pi} \int_{\bx} W(\bx) W(\bx+\br) \mathcal{L}_\ell(\hbr\cdot\hbd),
    \label{eq:windowQ_ell}
\end{align}
where $\hbd$ is the direction toward the pair $(\bx,\bx+\br)$.

After the calculation shown in Appendix~\ref{sec:window_convolution}, we finally obtain the unified expression for window convolutions for various power spectra:
\begin{align}
    \tilde{P}^{(\ell)}_{\rm X}(k) 
    = 4\pi(-i)^\ell \int r^2 \mathrm{d}r j_{\ell}(kr) \sum_{\ell'} Q^{\rm X}_{\ell \ell'}(r) 
    \left[ i^{\ell'} \int \frac{k'^2 \mathrm{d}k'}{2\pi^2} j_{\ell'}(k'r) P^{(\ell')}_{\rm X}(k')  \right],
    \label{eq:window_convolution_X}
\end{align}
where 
\begin{align}
    Q^{\rm X}_{\ell \ell'}(r) &\equiv \sum_{\ell''} Q_{\ell''}(r) (2\ell+1) \sqrt{\frac{(\ell-m_{\rm X})!}{(\ell+m_{\rm X})!} \frac{(\ell'+m_{\rm X})!}{(\ell'-m_{\rm X})!}}
    \begin{pmatrix}
        \ell'' & \ell & \ell'\\
        0 & 0 & 0
    \end{pmatrix}
    \begin{pmatrix}
        \ell'' & \ell & \ell'\\
        0 & m_{\rm X} & -m_{\rm X}
    \end{pmatrix},
    \label{eq:def_of_Q_llp}
\end{align}
with $({\rm X},m_{\rm X}) \in \left\{ (\delta\delta,0), (\gamma\delta,2), (+,0), (-,4) \right\}$. 
The $2\times3$ matrix form represents the Wigner $3j$ symbol. 
Note that the clustering case, $X=\delta\delta$, corresponds to the results shown in Ref.~\citep{Beutler&McDonald2021:wide_angle}\footnote{We omit the integral constraints and the wide-angle corrections for simplicity.}. 
In the case of ${\rm X} \in \left\{\gamma\delta,-\right\}$, $\ell$ and $\ell'$ in Eq.~(\ref{eq:window_convolution_X}) and (\ref{eq:def_of_Q_llp}) have to be considered as $L$ and $L'$, which means the labels of the associated Legendre polynomials.
Comparing the definition of the Hankel ($\mathcal{H}$) and inverse Hankel ($\mathcal{H}^{-1}$) transformations, we can reexpress Eq.~(\ref{eq:window_convolution_X}) as
\begin{align}
    \tilde{P}^{(\ell)}_{\rm X}(k) 
    = \mathcal{H}_\ell \left[ \sum_{\ell'} Q^{\rm X}_{\ell \ell'}(r) \mathcal{H}^{-1}_{\ell'} \left[ P^{(\ell')}_{\rm X}(k') \right](r) \right](k)
    = \mathcal{H}_\ell \left[ \sum_{\ell'} Q^{\rm X}_{\ell \ell'}(r) \xi^{(\ell')}_{\rm X}(r) \right](k).
    \label{eq:window_convolution_X_hankel}
\end{align}
Thus once we calculate the window correlation function multipoles, $Q_\ell$, by the pair-counting of  random particles that trace the survey geometry, the convolution effects on the measured power spectrum multipoles are now just a sum of the window multiplications between the inverse Hankel and Hankel transformations, which can be evaluated by 1D FFTs known as FFTlog \citep{Hamilton2000:FFTlog}.
This fast and precise convolution scheme was originally proposed by Ref.~\citep{Wilson+2017} for the galaxy clustering power spectrum and we here find that the same method can be applied to the case of the projected tensor power spectra.
Although the above formulae are general, we will hereafter assume $W(\bx)=1$ in the measured region and $0$ in the unmeasured region, respectively, and ignore the small-scale effects, e.g. masks due to bright stars.

\section{Validation Test}
\label{sec:validation}

In this section, we show a validation of the LPP power spectrum estimators we developed in the preceding section, assuming a survey geometry that mimics the \rvred{BOSS-like} survey footprint. 
We divide the validation test into two parts: first we demonstrate how well the LPP estimator works in curved sky configuration, compared with the GPP estimator. 
Then we compare the measurements from the LPP estimator with the theoretical prediction including the survey window convolution.

\subsection{Data and Settings}
\label{subsec:simulation_data}

To perform our validation test we use 1000 simulation realizations of the tidal field in a BOSS-like survey footprint, as described in the following.
We first generate each \rvred{realization} of the matter density field, $\delta(\bk)$ using the linear matter power spectrum $P(k)$ at redshift $z=0$, in a simulation box with comoving side length of $3~h^{-1}{\rm Gpc}$ with $512^3$ grids.
The Nyquist frequency $k_{\rm Ny} \simeq 0.5 ~h{\rm Mpc}^{-1}$. 
As for the input $P(k)$ we assume the flat-geometry $\Lambda$CDM cosmology, which is consistent with the {\it Planck} CMB data \citep{Planck2015_cosmo}: 
$\Omega_{\rm m}=0.3156$ for the matter density parameter, 
$\omega_{\rm b}(\equiv \Omega_{\rm b}h^2)=0.02225$, 
$\omega_{\rm c}(\equiv \Omega_{\rm c}h^2)=0.1198$ for the physical density parameters of baryon and CDM, 
and $n_{\rm s}=0.9645$ and $\ln(10^{10}A_{\rm s})=3.094$ for the tilt and amplitude parameters of the primordial curvature power spectrum. 
The input power spectrum corresponds to $\sigma_8=0.834$, the rms value of \rvred{present-day} mass fluctuations within a sphere of radius $8\,h^{-1}{\rm Mpc}$.
We then compute the tidal field $T_{ij}(\bk) \equiv (\hk_i \hk_j - \delta_{ij}/3) \delta(\bk)$ in Fourier space  
and inverse-Fourier-transform the field to obtain $T_{ij}(\bx)$ on each grid in configuration space.
We repeated the above procedures to generate 1000 realizations of $T_{ij}$ using different random seeds. 
In this validation test, we simply assume that the tidal field itself is inferred from some tensor-type large-scale structure observables such as galaxy shapes. 
Although this is a very simplified setup, we think this is enough to validate both our LPP estimators and window convolution calculations for the following reasons.
Our main interest is to study how well the LPP estimator recovers the desired signal at large scales where the GPP approximation ceases to be accurate.
For the BOSS-like survey geometry, the window effects become significant at large scales $k\lesssim 0.1\,h{\rm Mpc}^{-1}$. 
In the standard structure formation model with GR, large-scale correlations of any traceless scalar component should arise from the tidal field, i.e. the longitudinal scalar mode or helicity-0 component defined in Eq.~(\ref{eq:helicity_decomposition_of_sij}). 
For example the large-scale IA signal of galaxy shapes is given by the linear alignment model \citep[e.g.][]{Hirata&Seljak2004:IA_LA}. 
Nonlinear structure formation generally induces other helicity components \citep[e.g.][]{Blazek+2017:IA_TATT,Vlah+2020:IA_EFT}, but these appear only at small scales where the LPP estimator should work and the window convolution effect is small. 
Therefore we consider that our setup is sufficient to validate our method. 

We further define the projected tensor field observed at each grid as 
\begin{align}
    \gamma(\bx) \equiv e_{ij}(\hbx_{\rm obs}) T_{ij}(\bx).
    \label{eq:validation_projection}
\end{align}
From Eqs.~(\ref{eq:def_of_prj_cross_power})--(\ref{eq:def_of_prj_auto_power_-}), the underlying power spectra of the projected tensor field are given as
\begin{align}
    P_{\gamma \delta}(\bk) &= \frac{1}{2} (1-\mu_k^2) P(k), \nonumber\\
    P_{+}(\bk) = P_{-}(\bk) &= \frac{1}{4} (1-\mu_k^2)^2 P(k).
    \label{eq:power_for_validation}
\end{align}
In the following we do not consider the RSD effect, so the $\mu_k$-dependence of the power spectra is from the geometrical factor, $(1-\mu_k^2)$ or $(1-\mu_k^2)^2$.

We employ a survey window that mimics the BOSS Northern Galactic cap (NGC) footprint \citep{2017MNRAS.470.2617A}. 
To be more precise, we place a hypothetical observer at one particular position in each simulation box and then cutout the fields that lie in the hypothetical survey region with angular ranges of ${\rm RA} \in [120^\circ,260^\circ]$ and ${\rm DEC} \in [-5^\circ,70^\circ]$ and the radial distance range of $r \in [1.0,1.5]~h^{-1}{\rm Gpc}$, as illustrated in Fig.~\ref{fig:los_configuration}. 
Here the radial window roughly corresponds to the redshift range $0.35\lesssim z\lesssim 0.55$ for the $\Lambda$CDM model or the redshift range of the ``mid-$z$ NGC sample'' used in the cosmology analysis \citep[e.g.][]{Beutler+2017}. 
The survey geometry has the solid-angle area of about $ 8,000$~square degrees on the celestial sphere for the observer and the comoving volume of about $2~(h^{-1}{\rm Gpc})^3$.

\subsection{Results: LPP vs GPP}
\label{subsec:lpp_vs_gpp}

\begin{figure*}
    \centering
    \includegraphics[width=1.0\columnwidth]{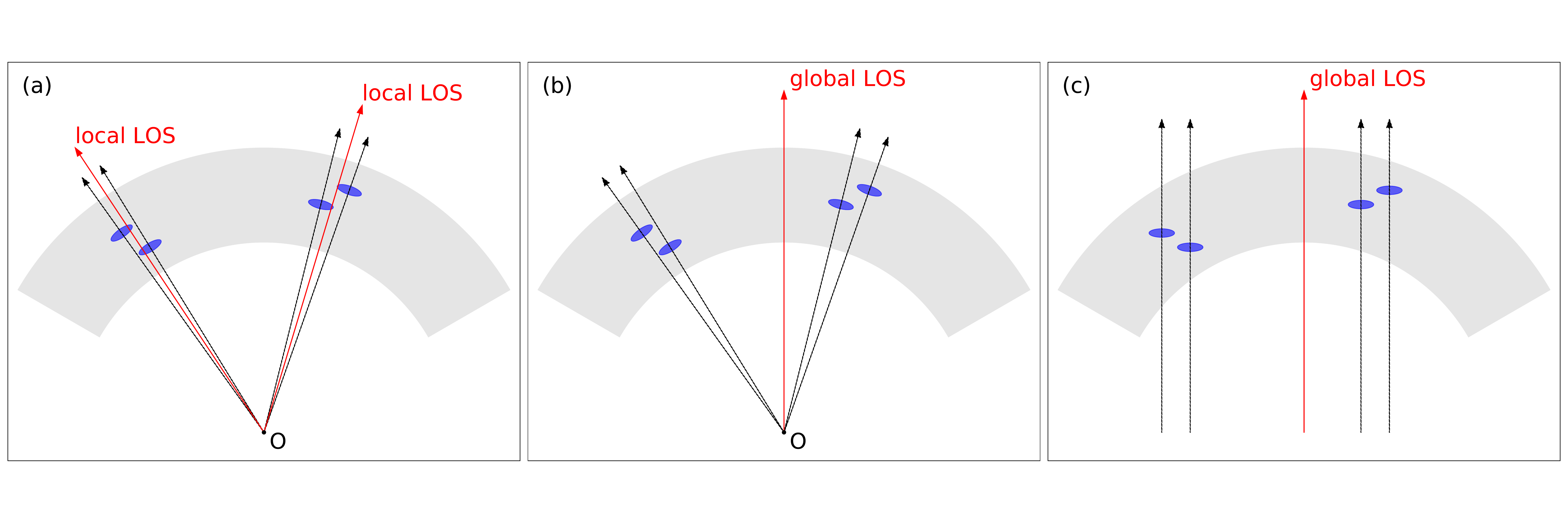}

    \caption{A schematic illustration of setups used for validation of the LPP power spectrum estimators for the projected tensor field.
    Gray shaded region in each panel denotes a hypothetical survey window for which we take the BOSS Northern Galactic cap (NGC) footprint; its solid angle area $\Omega_{\rm sky}\simeq 8,000~$deg$^2$ and the radial width $\Delta r=500\,h^{-1}{\rm Mpc}$ (see text for details). 
    Each blue ellipse denotes a ``shape'' of the projected tidal field at its position; we define the field by projecting the 3D tidal field onto plane perpendicular to the LOS direction, denoted by the black arrow. 
    {\it Case~(a)}: This setup is for validating the LPP estimators. 
    For each pair taken in the power spectrum estimate, we use the local LOS direction, denoted as red arrow, to compute the phase factor and the associated Legendre polynomials, which are needed for estimation of the multipole moments (see text for details). 
    {\it Case~(b)}: The definition of the projected tidal field is the same as Case~(a), but we use the GPP estimator to measure the multipole moments of power spectra. 
    The global LOS direction (red arrow) is set as the direction of the midpoint of the survey region, viewed by the observer position.
    {\it Case~(c)}: This is the case where we can employ the distant observer approximation. 
    In this case we can use the global LOS direction to define the projected tidal field and to implement the GPP estimator. 
    We use the same LOS direction as the red arrow in Case~(b). 
    To have a fair comparison, we employ the same survey window as Case~(a) and (b). 
    For all the three cases we use exactly the same simulation realizations of the underlying 3D tidal field.
    }
    \label{fig:los_configuration}
\end{figure*}

In this section we compare the power spectra that are measured from the fields $\{ \delta(\bx), \gamma(\bx) \}$ in each realization using the GPP or LPP estimators.
To validate or assess the accuracy of the LPP estimator for the projected tensor field, we consider the following three configurations of the projection, as schematically illustrated in Fig.~\ref{fig:los_configuration}:
\begin{itemize}
    \item Case (a): This is the setup that mimics an actual observation, intended to validate
    the LPP estimators developed in this paper.
    We put an observer on the origin and define the LOS direction to each grid point $\bx$, i.e. $\hbx_{\rm obs} = \hbx$. We define the projected tensor field from the proper projection by Eq.~(\ref{eq:validation_projection}): $\gamma(\bx) \equiv e_{ij}(\hbx)T_{ij}(\bx)$. 
    Then we measure the power spectra of the projected field using the LPP estimator (Eq.~\ref{eq:lpp_estimator_cross_Xiij}). 
    \item Case (b): This setup is for evaluating the inaccuracy of GPP estimator in the power spectrum measurement. 
    For the projected tensor field, we use the same definition as in Case~(a): $\hbx_{\rm obs} = \hbx$ at each grid position. 
    However, for the power spectrum measurement, we adopt the global LOS direction ($\hbn$) toward the midpoint of the survey region to implement the GPP estimator; 
    we set $\hbd \rightarrow \hbn$ in Eq.~(\ref{eq:lpp_estimator_cross})\footnote{Equivalently we set $e^\ast_{ij}(\hbx)/[1-(\hbk\cdot\hbx)^2] \rightarrow e^\ast_{ij}(\hbn)/[1-(\hbk\cdot\hbn)^2]$ for the phase factor and ${\cal L}^{m=2}_L(\hbk\cdot\hbx)\rightarrow {\cal L}^{m=2}_L(\hbk\cdot\hbn)$ for the associated Legendre polynomials, respectively, in Eq.~(\ref{eq:lpp_estimator_cross_Xiij}).} 
    for the cross power spectrum estimator and thus the estimator become
    \begin{align}
        \rvred{\hat{P}}^{(L)}_{\gamma \delta}(k) 
        &\rightarrow (2L+1) \frac{(L-2)!}{(L+2)!} \int_{\hbk, \bx,\bx'} \gamma(\bx) \delta(\bx') e^{-2i\phi_{\hbk,\hbn}} e^{-i\bk \cdot (\bx - \bx')} \mathcal{L}^{m=2}_{L}(\hbk \cdot \hbn) ~\nonumber\\
        &= (2L+1) \frac{(L-2)!}{(L+2)!} \int_{\hbk} [E(\bk;\hbn)+iB(\bk;\hbn)] \delta(-\bk) \mathcal{L}^{m=2}_{L}(\hbk \cdot \hbn),
        \label{eq:gpp_estimator_cross}
    \end{align}
    which is equivalent to the one used in Ref.~\citep{Kurita+2020:IA_nbody}. 
    Therefore this case uses an inconsistent treatment for the projection for the tensor field and for the power spectrum estimator. 
    \item Case (c): This is the case for distant observer approximation. 
    Namely we set $\hbx_{\rm obs} \equiv \hbn$ to define the projected tensor field; $\gamma(\bx) \equiv e_{ij}(\hbn)T_{ij}(\bx)$ and implement the GPP estimator (Eq.~\ref{eq:gpp_estimator_cross}) for the power spectrum measurement. 
    To have a fair comparison, we use the same survey window as Case~(a) and (b). 
\end{itemize}
We use the cubic volume of $27\,(h^{-1}{\rm Gpc})^3$, which is the same as that of the original simulations, to perform the FFT transform. 
To reduce the statistical errors, in the following we show the average of the power spectra measured from the 1000 realizations.

\begin{figure*}
    \centering
    \includegraphics[width=1.0\columnwidth]{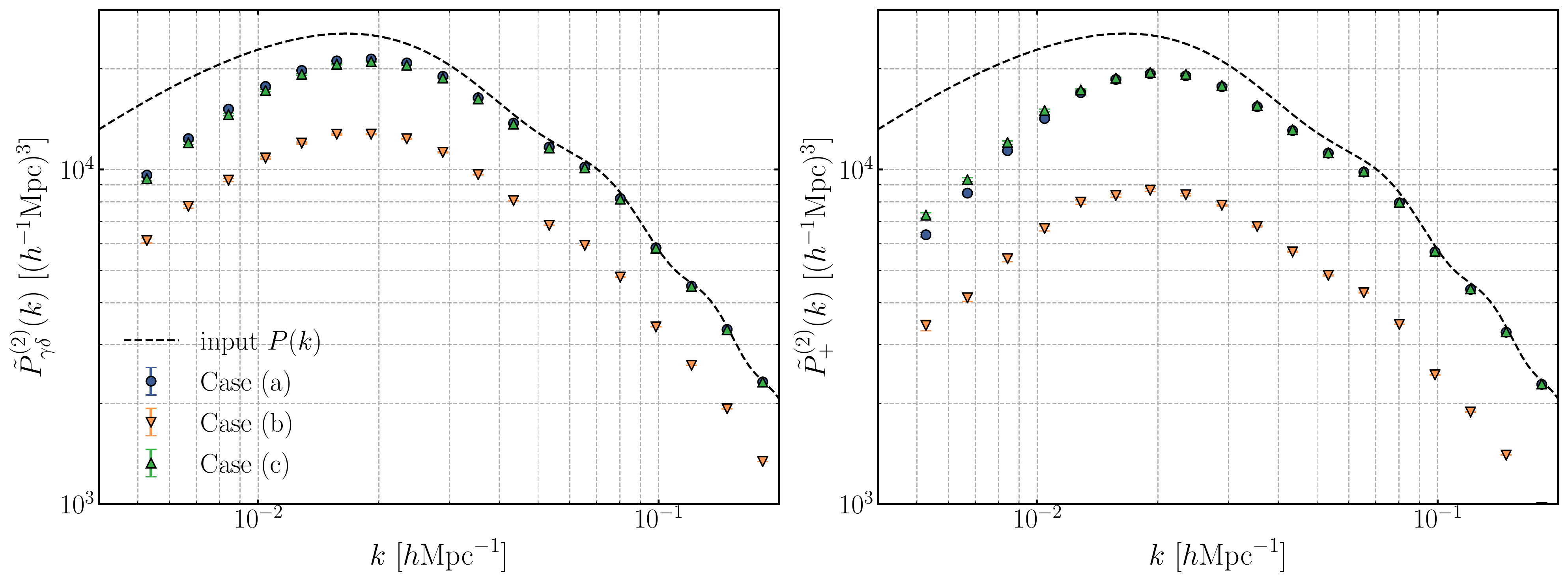}
    \caption{{\it Left panel}: The lowest-order multipole moment of the cross-power spectrum of the projected tidal field, $P^{(2)}_{\gamma\delta}$, measured using the LPP or GPP estimators in each of the three setups illustrated in Fig.~\ref{fig:los_configuration}.
    The blue circle symbols show the result for Case~(a), i.e. from the LPP estimator developed in this paper, while the green-triangle symbols are the result for Case~(c), the GPP estimator under the distant observer approximation. 
    The orange inverted-triangle symbols are the result for Case~(b), which uses the inconsistent LOS directions for the projected tensor field and for the power spectrum estimator. 
    The dashed line denotes the input matter power spectrum from which the tidal field is simulated. 
    {\it Right panel}: Similar to the left panel, but the results for the quadrupole moment of the auto-power spectrum of the projected tidal field, $P^{(2)}_+(k)$.
    To make the comparison easier, we multiply all the results by normalization factor to match them with the input $P(k)$ amplitude (see text for details). 
    }
    \label{fig:lpp_vs_gpp}
\end{figure*}

Fig.~\ref{fig:lpp_vs_gpp} compares the multipole moments of the power spectra of the projected tidal field, $\tilde{P}^{(L)}_{\gamma\delta}$ and $\tilde{P}_+^{(\ell)}$ with $L=2$ and $\ell=2$, respectively, measured using the LPP or GPP estimators in Case~(a)--(c) in Fig.~\ref{fig:los_configuration}. 
As shown in Eq.~(\ref{eq:power_for_validation}), the multipole moments of the underlying power spectra, $P_{\gamma\delta}$ and $P_+$, differ from the matter power spectrum $P(k)$ by a constant factor that is from the $\mu_k$-integral of the geometrical factor, $(1-\mu_k^2)^n$ ($n=1$ or 2), weighted by the respective basis at each order given in Eq.~(\ref{eq:xi_relations_normalization_coefficients}).
We multiply the simulated results by this factor so that the results match $P(k)$ if the window function is absent, for illustrative purpose.
First of all, the results for Case~(a) and (c) fairly well agree with each other, verifying that the LPP estimator developed in this paper works. 
However, note that the agreement is not necessarily perfect, and we expect subtle difference because the window functions $Q_\ell$ in Case~(a) and (c) are different with each other due to the different LOS assumptions (see Eq.~{\ref{eq:windowQ_ell}}) and hence the resulting window convolution arises from different Fourier modes in the two cases.
Both Case~(a) and (c) results show that the BOSS-like survey window (Fig.~\ref{fig:los_configuration}) affects the mulitpoles at $k\lesssim 0.1\,h{\rm Mpc}^{-1}$; the window effect leads to scale-dependent suppression in the multipole amplitudes at the scales compared to the input $P(k)$.
On the other hand, the Case~(b) result displays a significant deviation from the input $P(k)$ or the Case~(a)/(c) result and also does not match 
the input even at small scales because the anisotropic signal from the galaxy pairs far from the global LOS direction, i.e. the midpoint vector of the survey geometry, is not averaged with the correct weighting due to the inconsistent LOS assumptions for the projected tensor field and for the estimation.
Note that the discrepancy of the Case~(b) from Case~(a) results is scale-dependent at $k\lesssim 0.1\,h{\rm Mpc}^{-1}$ and therefore cannot be absorbed by a change in the normalization factor. 
Thus we conclude that the LPP estimator works well to have an almost unbiased measurement of the power spectrum of the projected tidal field.

\subsection{Results: LPP Estimator vs Theory Prediction}
\label{subsec:lpp_vs_theory}

\begin{figure*}
    \centering
    \includegraphics[width=0.7\columnwidth]{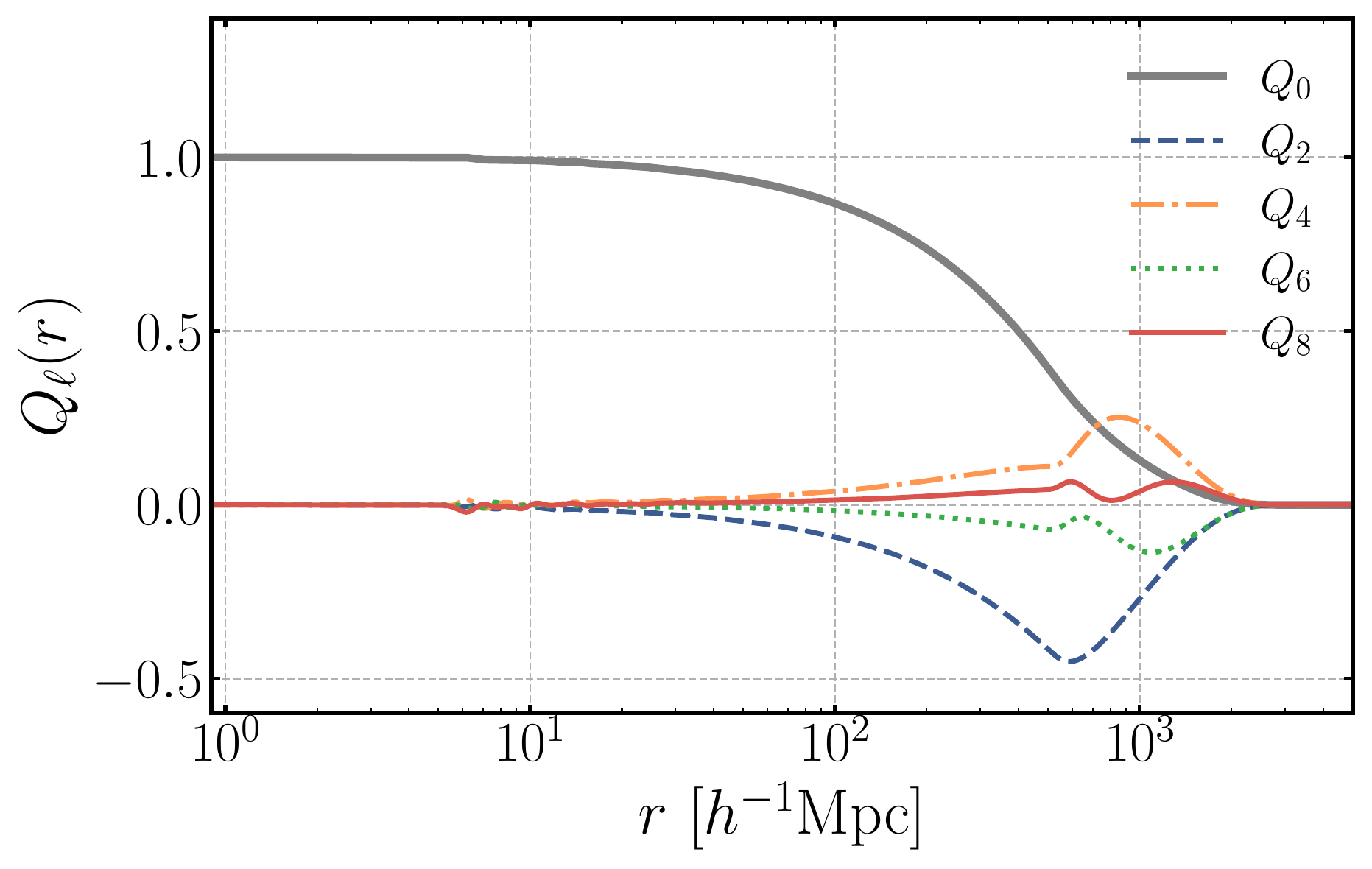}
    \caption{Multipole moments of the auto-correlation function of survey window, $Q_\ell$ (Eq.~\ref{eq:windowQ_ell}), computed using the pair-counting method for the BOSS-like survey window in Fig.~\ref{fig:los_configuration}.
    }
    \label{fig:window_qfunctions}
\end{figure*}
\begin{figure*}
    \centering
    \includegraphics[width=1.0\columnwidth]{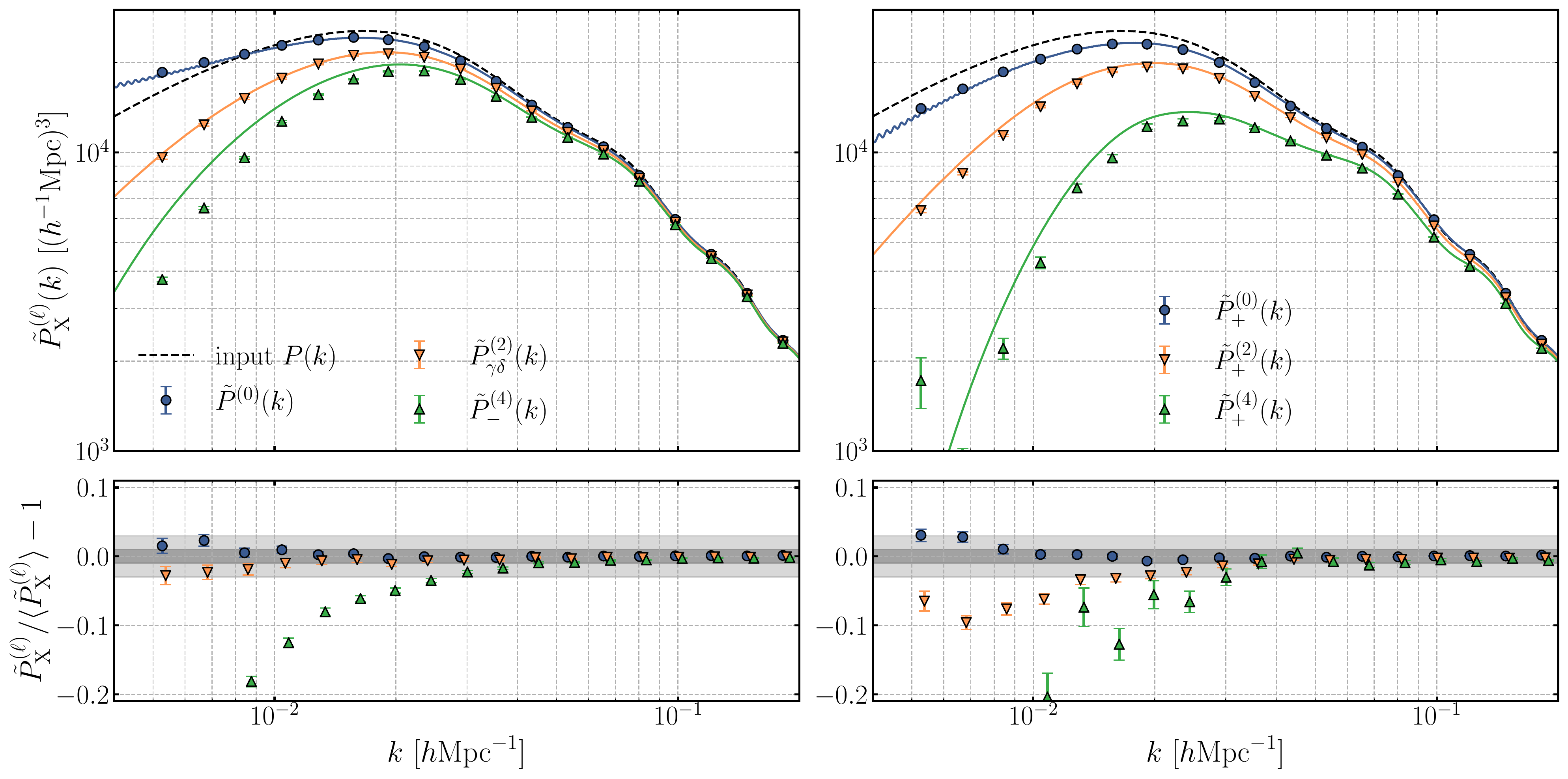}
    \caption{The upper panels compare the multipole moments of power spectra, measured by the LPP estimator (points with errorbars), with the theoretical predictions including the window effects (solid lines). 
    The lower panels show the fractional differences. 
    We plot the mean of 1000 realizations with errorbar of the mean in each $k$ bin, and  
    normalize all the moments as in Fig.~\ref{fig:lpp_vs_gpp}.
    {\it Left panel}: The blue circles, orange inverted triangles and green triangles show the matter spectrum $\tilde{P}^{(0)}$, the lowest-order $L=2$ moment of cross spectrum $\tilde{P}^{(L=2)}_{\gamma\delta}$ and the lowest-order $L=4$ moment of ``minus''-auto spectrum $\tilde{P}^{(L=4)}_{-}$, respectively, which are the lowest order multipole moments for each spectrum. 
    {\it Right}: The blue circles, orange inverted triangles and green triangles show the $\ell=0,2,4$-th moments of the ``plus''-auto spectrum $\tilde{P}^{(\ell)}_+$, respectively. 
    The dark and light gray-shaded regions in the lower panel correspond to 1\% and 3\% fractional differences.
    }
    \label{fig:lpp_vs_theory_main}
\end{figure*}
\begin{figure*}
    \centering
    \includegraphics[width=0.5\columnwidth]{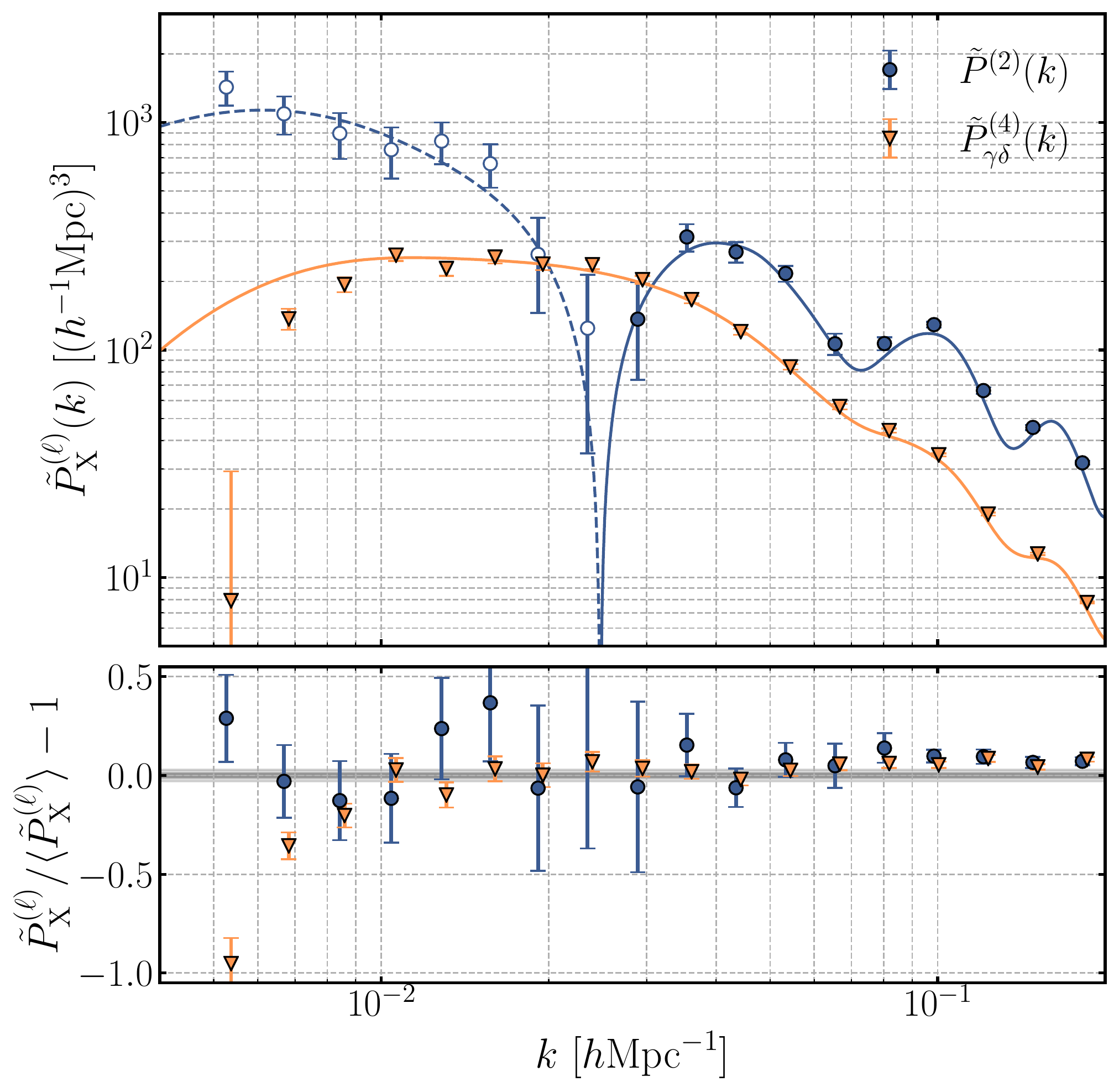}
    \caption{Similar to Fig.~\ref{fig:lpp_vs_theory_main}, but this plot shows the leakage moments that arises due to the 
    anisotropy of BOSS-like survey window; in other words, the moments should be vanishing if there is no window effect 
    or for an isotropic window.
    The blue circles and orange inverted triangles are the quadrupole moment of the matter spectrum $\tilde{P}^{(2)}$ and the $L=4$-th moment of the cross spectrum $\tilde{P}^{(L=4)}_{\gamma\delta}$, respectively.
    The blue open and blue dashed lines denote the negative values (we plot their absolute values). 
    Note that the range of $y$-axis is different from that in Fig.~\ref{fig:lpp_vs_theory_main}. 
    }
    \label{fig:lpp_vs_theory_leakage}
\end{figure*}
To interpret the measured power spectra of the projected tensor field such as those in Fig.~\ref{fig:lpp_vs_gpp}, we need theoretical templates of the multipole moments including the survey window effects. 
As we derived by Eq.~(\ref{eq:window_convolution_X_hankel}), we can compute the theoretical templates for the window-convolved power spectrum once the underlying power spectra of the projected tensor field and the correlation function of the window function, $Q(\br)$ (Eq.~\ref{eq:windowQ}), are given. 
In this section we evaluate whether the theoretical templates, computed based on our formula, can reproduce the multipole moments of power spectra measured from the simulations of BOSS survey footprint in Section~\ref{subsec:lpp_vs_gpp}. 

In our validation simulations of the tidal field, the underlying power spectra are given by Eq.~(\ref{eq:power_for_validation}); the power spectra, apart from the geometrical $\mu_k$-dependence of $(1-\mu_k^2)^n$, has no $\mu_k$-\rvred{dependence}, i.e. the real-space matter power spectrum, $P(k)$. 
In this case the moments, $P^{(L=2)}_{\gamma\delta}(k)$, $P_+^{(\ell=0,2,4)}(k)$ and $P_-^{(L=4)}(k)$, contain the full information.
Using the formula Eq.~(\ref{eq:window_convolution_X_hankel}), we can derive the theoretical predictions for these moments including the survey window effects: 
\begin{align}
    \tilde{P}^{(\ell=0)}(k) &= 4\pi\int_0^{\infty} r^2\mathrm{d}r j_0(kr) \xi^{(0)}(r) Q_0(r), \nonumber\\
    \tilde{P}^{(L=2)}_{E\delta}(k) &= -4\pi\int_0^{\infty} r^2\mathrm{d}r j_2(kr) \frac{1}{6} \xi^{(2)}(r) \left\{Q_0(r) -\frac{2}{7}Q_2(r) + \frac{1}{21}Q_4(r) \right\}, \nonumber\\
    \tilde{P}^{(\ell=0)}_{+}(k) &= 4\pi\int_0^{\infty} r^2\mathrm{d}r j_0(kr) 
    \left[ \frac{2}{15}\xi^{(0)}(r) Q_0(r) - \frac{4}{105}\xi^{(2)}(r) Q_2(r) + \frac{2}{315}\xi^{(4)}(r) Q_4(r)
    \right], \nonumber\\
    \tilde{P}^{(\ell=2)}_{+}(k) &= -4\pi\int_0^{\infty} r^2\mathrm{d}r j_2(kr) 
    \left[ 
    \frac{2}{15} \xi^{(0)}(r) Q_2(r) 
    - \frac{4}{21} \xi^{(2)}(r) \left\{ Q_0(r) + \frac{2}{7}Q_2(r) + \frac{2}{7}Q_4(r) \right\} \right. \nonumber\\
    &\left. \quad \quad \quad \quad \quad \quad \quad \quad \quad \quad + \frac{2}{35} \xi^{(4)}(r) \left\{\frac{2}{7}Q_2(r) + \frac{100}{693}Q_4(r) + \frac{25}{143}Q_6(r) \right\}
    \right], \nonumber\\
    \tilde{P}^{(\ell=4)}_{+}(k) &= 4\pi\int_0^{\infty} r^2\mathrm{d}r j_4(kr) 
    \left[ 
    \frac{2}{15} \xi^{(0)}(r) Q_4(r) 
    - \frac{4}{21} \xi^{(2)}(r) \left\{ \frac{18}{35}Q_2(r) + \frac{20}{77}Q_4(r) + \frac{45}{143}Q_6(r) \right\} \right. \nonumber\\
    &\left. \quad \quad \quad \quad \quad \quad \quad \quad \quad + \frac{2}{35} \xi^{(4)}(r) \left\{Q_0(r) + \frac{20}{77}Q_2(r) + \frac{162}{1001}Q_4(r) + \frac{20}{143}Q_6(r) + \frac{490}{2431}Q_8(r) \right\}
    \right], \nonumber\\
    \tilde{P}^{(L=4)}_{-}(k) &= 4\pi\int_0^{\infty} r^2\mathrm{d}r j_4(kr) \frac{1}{420} \xi^{(4)}(r) \left\{Q_0(r) -\frac{4}{11}Q_2(r) + \frac{18}{143}Q_4(r) -\frac{4}{143}Q_6(r) + \frac{7}{2431}Q_8(r)\right\}, 
    \label{eq:all_convolutions}
\end{align}
where $\xi^{(\ell)}$ is the $\ell$th order inverse Hankel transform of the input matter power spectrum:
\begin{align}
    \xi^{(\ell)}(r) \equiv i^\ell \int_0^\infty \frac{k^2 \mathrm{d}k}{2\pi^2} P(k) j_\ell(kr) = \mathcal{H}^{-1}_\ell[P(k)](r),
\end{align}
and we have used the relations:
\begin{align}
    \xi^{(L=2)}_{\gamma\delta} = \frac{1}{6}\xi^{(2)},~ \left\{ \xi^{(\ell=0)}_{+}, \xi^{(\ell=2)}_{+}, \xi^{(\ell=4)}_{+}\right\} = \left\{ \frac{2}{15}\xi^{(0)}, -\frac{4}{21}\xi^{(2)}, \frac{2}{35}\xi^{(4)}\right\},~ \xi^{(L=4)}_{-} = \frac{1}{420}\xi^{(4)}.
    \label{eq:xi_relations_normalization_coefficients}
\end{align}
Here we also gave the monopole moment of the input matter power spectrum, $\tilde{P}^{(0)}$, for comparison. 

First, Fig.~\ref{fig:window_qfunctions} shows the multipole moments of $Q_\ell(r)$ (Eq.~\ref{eq:windowQ_ell}), which is estimated by the pair-counting method of random points generated in the survey footprint \citep{Wilson+2017}. 
From Eq.~(\ref{eq:all_convolutions}) we show the results up to $Q_8$. 
The figure shows that the BOSS-like survey window yields the higher-order multipoles, although the higher-order moments have smaller amplitudes than the lower-order ones. 

Fig.~\ref{fig:lpp_vs_theory_main} shows another main result of this paper. 
The figure shows that the measurement and prediction for the lowest order moments, $\tilde{P}^{(2)}_{\gamma\delta}$ and $\tilde{P}^{(0)}_{+}$, are in good agreement with each other to within 1\% in the amplitude 
\rvred{down}
to $k \simeq 0.01~h{\rm Mpc}^{-1}$.
Note that the errorbars in each $k$ bin are $1\sigma$ errors on the mean, and the statistical error expected for the BOSS-like survey is about factor of 30 greater than the errors shown here. 
The agreement indicates that the LPP estimator works well, and is ready to use for  cosmological analysis.
On the other hand, the minus auto-power spectrum, $\tilde{P}_-^{(4)}$, and the higher-order moments of $\tilde{P}_+$ such as $\tilde{P}^{(2)}_{+} $ and $\tilde{P}^{(4)}_{+}$, display more than 3\% deviations from the theoretical predictions, with increasing deviations in lower $k$ bins at typically $k \lesssim 0.03~h{\rm Mpc}^{-1}$. 
We think that the deviations are due to 
\rvred{breakdown of the plane-parallel approximation}
and can be explained by the wide-angle effect \citep[e.g.][for such discussion on the \rvred{density} power spectrum]{Castorina&White2018:wide_angle} (also see \citep{Shiraishi+2021:IA_WA}). 
This is beyond the scope of this paper, and will be our future work.
Another important remark is that the survey window effect gives more significant suppression in the moment amplitudes of the power spectrum for the projected tidal field, compared to that for the standard power spectrum of the density field ($\tilde{P}^{(\ell=0)}$ here). 
This is mainly because the projection prefactor, $(1-\mu_k^2)^n$, makes the underlying power spectrum more anisotropic and thus causes a more coupling of the Fourier modes with the higher moments of the window function, $Q_\ell$.
This means that the window effect is important to take into account in order to make a correct interpretation of the measured power spectra of the projected tensor field. 

For comprehensiveness of discussion,  we also study the moments that arise from leakage of the power due to the survey window effect: 
\begin{align}
    \tilde{P}^{(\ell=2)}(k) &= -4\pi\int_0^{\infty} r^2\mathrm{d}r j_2(kr) \xi^{(0)}(r) Q_2(r), \nonumber\\
    \tilde{P}^{(L=4)}_{E\delta}(k) &= 4\pi\int_0^{\infty} r^2\mathrm{d}r j_4(kr) \frac{1}{6} \xi^{(2)}(r) \left\{\frac{3}{35}Q_2(r) -\frac{6}{77}Q_4(r) + \frac{3}{143}Q_6(r) \right\}.
\end{align}
Note that these moments should be vanishing if there is no survey window effect. 
Fig.~\ref{fig:lpp_vs_theory_leakage} shows the leakage moments have a few per cent powers compared to the main signals, $\tilde{P}^{(\ell=0)}$ and $\tilde{P}^{(L=2)}_{\gamma\delta}$.
We also show that the above theoretical prediction, based on our formula, can well reproduce 
the leakage moments.

\section{Discussion}
\label{sec:discussion}
\begin{figure*}
    \centering
    \includegraphics[width=0.75\columnwidth]{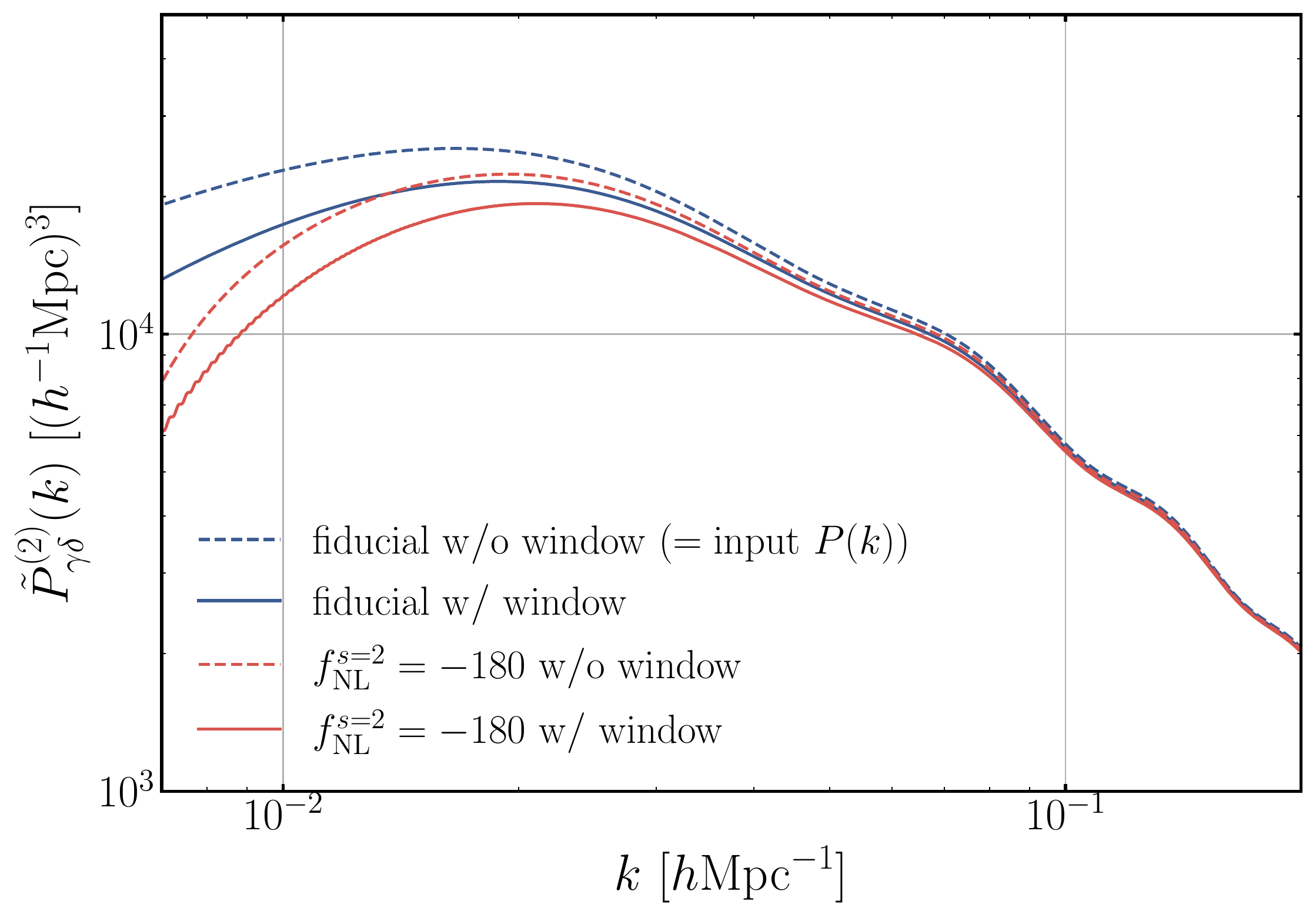}
    \caption{
    Comparison of the BOSS-like survey window effect on the cross power spectrum $P_{\gamma\delta}^{(L=2)}$ with the modification in $P^{(L=2)}_{\gamma\delta}$ due to the spin-2 local-type primordial non-Gaussian (PNG) initial condition. 
    Here we show the quadrupole moment of $P_{\gamma\delta}$ because it is the lowest-order moment in our formulation and carries most of the PNG information. 
    The blue curves show the power spectra in the Gaussian condition and the red curves are those in the presence of the spin-2 PNG.
    The solid (dashed) lines correspond  to the results with (without) the window convolution.
    We chose $f^{s=2}_{\rm NL}=-180$ for demonstration, because the effect without the window convolution (dashed-red curve) is at the comparable level with the Gaussian result with the window convolution (solid-blue curve). 
    }
    \label{fig:fnl_discussion}
\end{figure*}

We would like to mention about several things looking ahead to application of our method to real data.
Direct observables of the auto power spectra are the ``plus-'' and ``minus-'' spectra, $P_+$ and $P_-$, not the spectra of the ``$E$'' and ``$B$'' modes as shown in Eq.~(\ref{eq:relation_pm_EB}). 
Although $\left\{P_+(k,\mu),P_-(k,\mu)\right\}$ and $\{P_{EE}(k,\mu),P_{BB}(k,\mu)\}$ are exchangeable at the level of functions of $(k,\mu)$, the direct observables in our method are the angle-integrated moments, $\left\{P^{(\ell=0,2,4,\cdots)}_+(k),P^{(L=4,6,\cdots)}_-(k)\right\}$, in terms of different basis, i.e. the Legendre and associated Legendre polynomials, respectively.
Hence, it is not straightforward to reconstruct the $E$- and $B$-mode spectra from the measured moments, where the $E$- and $B$-mode spectra might be used to measure the power spectrum of the dominated scalar mode for the $\Lambda$CDM model and to test residual systematics in the linear regime, respectively, e.g. as expected in the case of the IA spectra and the cosmic shear angular power spectra.
Nevertheless, we stress that a set of the multipole moments of $P_+$ and $P_-$ can be used to extract the full information on the underlying power spectra of scalar, vector and tensor modes according to our method, as shown in Section \ref{subsec:connection_with_correlation_function}.
In addition, if a stochastic noise such as shape noise of galaxies equally contributes to the $E$- and $B$-mode spectra, only the monopole of $P_+$ has the shape noise contribution, and other higher-order moments of $P_+$ and all moments of $P_-$ can be free of the stochastic noise. In any case we can utilize these properties to perform tests of systematic effects in an actual measurement. 

An obvious application of our method is to explore the spin-2, local-type anisotropic primordial non-Gaussianity (PNG) from the power spectrum of the projected tensor field, which carries an independent information from the isotropic PNG in the density tracer \citep{Akitsu+2021:IA_PNG}. 
Note that, to explore the PNG signal, the power spectrum analysis is needed, and the correlation function in configuration space is not suitable \citep{2008PhRvD..77l3514D}.
In the presence of spin-2 PNG, the power spectrum, probed by galaxy shapes, is modified as 
\begin{align}
    P(k) \to \left[ 1+12f^{s=2}_{\rm NL} \delta b_{\rm IA} \mathcal{M}^{-1}(k) \right]P(k)
\end{align}
where $\mathcal{M}(k,z) \equiv (2/3)k^2T(k)D(z) / (\Omega_{\rm m}H^2_0)$, with $T(k)$ and $D(z)$ denoting the transfer function and the linear growth factor, respectively, $f^{s=2}_{\rm NL}$ is a parameter to characterize the PNG amplitude and $\delta b_{\rm IA}$ is the response of galaxy shapes to the quadrupolar modulation of the local matter power spectrum due to the PNG. 
In the following we adopt $\delta b_{\rm IA} = 0.17$ as motivated by the results in Ref.~\citep{Akitsu+2021:IA_PNG}.

Fig.~\ref{fig:fnl_discussion} shows how the window convolution affects the quadrupole moments of the cross power spectrum, $P^{(L=2)}_{\gamma\delta}$, which carries a leading signal of the PNG effect \citep{Akitsu+2021:IA_PNG}. 
It is clear that the survey window effect for the BOSS-like survey is significant. 
For comparison, we also show that the window effect alters the power spectrum with the Gaussian initial condition (i.e. the case of $f_{\rm NL}^{s=2}=0$) at the comparable level to the modification in the power spectrum with $f_{\rm NL}^{s=2}\sim -180$ in the absence of the survey window effect.
Since the current CMB constraint is at the level of $\sigma(f_{\rm NL}^{s=2})\sim 20$ \citep{2020A&A...641A...9P}, it is of critical importance to take into account the survey window effect, using our method, to obtain an unbiased estimate or constraint of the PNG signal in $P_{\gamma\delta}$.

\section{Conclusion}
\label{sec:conclusion}

In this paper, we have developed a method for measuring the 3D power spectrum of the projected tensor field that is estimated from large-scale structure observables, e.g. galaxy shapes. 
The projected tensor power spectrum is related to the underlying power spectrum of scalar, vector and tensor modes of LSS, so the measurement opens up a window for exploring these different types of perturbations. 

In a wide-area survey where the global plane-parallel (GPP) approximation (or the distant observer approximation) is no longer valid, the statistical translation invariance does not hold for the observed tensor field due to the {\it LOS dependent} projection, very similarly to the case of the observed galaxy density field affected by the {\it LOS dependent} RSD effect. 
To obtain both the estimators and window convolutions, we first formulated the coordinate-independent power spectra and correlation functions taking into account the projection of the tensor field to plane perpendicular to each LOS ($\hbn$) direction that leads to the phase factor ($e^{2i\phi_{\hbk,\hbn}}$ in Eq.~\ref{eq:def_of_phase_factor}) reflecting the spin-2 properties of the tensor field.
In addition we derived the Hankel transformation between the multipole moments of the power spectra and those of the correlation functions by introducing the associated Legendre polynomial expansion. 
The expansion in terms of the associated Legendre polynomials provides us with two crucial results for the analysis proposed in this paper: 
\begin{itemize}
    \item For measurements of the projected tensor power spectra, the exact cancellation of the geometrical prefactors allows us to construct FFT-based estimators under the LPP approximation even in a realistic wide-area survey.
    \item The Hankel transformation relation between the two-point statistics in Fourier and configuration space allows us to calculate the survey window effects in the theoretical prediction for the multipole moments of projected tensor power spectrum based on the 1D Fourier transforms (FFTlog).
\end{itemize}

To validate our formulations, we used the simulations of tidal field to perform a hypothetical measurement of the projected tensor power spectrum assuming the BOSS NGC-like survey geometry. 
We simply considered the tidal field, simulated assuming the random Gaussian field, as the underlying tensor field and, using this, defined the observed tensor field performing the projection on the normal plane to the LOS direction of each pair in the power spectrum estimation.
This mimics a local observable linearly aligned with the surrounding tidal field, which is the case for intrinsic alignments of galaxy shapes at large scales for instance.
Although this seems to be a simplified setup, we believe that this is enough for us to validate our method because our main interest is how well our LPP estimators work at large scales where the GPP approximation breaks down. 
From the validation test, we found that the measurements and the theoretical window convolutions of the lowest moments of the cross- and plus- power spectrum, $P^{(2)}_{\gamma\delta}$ and $P^{(0)}_{+}$ are consistent with each other at 1\% level even at $k\simeq0.01~h{\rm Mpc}^{-1}$ that is the minimum wavenumber accessible from the BOSS-like survey. 
On the other hand, for the higher order moments such as $P^{(2,4)}_{+}, P^{(4)}_{-}$, there are more than 3\% discrepancies between the measurement and theory at $k \lesssim 0.03~h{\rm Mpc}^{-1}$, probably due to the wide-angle effect. 
Hence if we want to use the higher-order moments in an analysis, we have to be careful about this possible bias. 
Even in this case, our method can quantify this possible bias for the $\Lambda$CDM-like linear power spectrum, for a given survey geometry. 
Alternatively the wide-angle correction can be included to further improve the accuracy of our method on the lower wavenumber scales.
We should also stress that our LPP based method would be more accurate for upcoming wide-area surveys such as the Subaru PFS and DESI, which probe large-scale structure at higher redshifts, where lower wavenumbers correspond to smaller angular separation in each galaxy pair than in a lower-redshift galaxy survey. 

With the analysis method in this paper we are ready to perform a cosmology analysis of the projected tensor power spectra
for actual data. 
An obvious direction is to explore the primordial non-Gaussianity (PNG) signal from the cross power spectrum of galaxy shapes and galaxy number density fluctuation on small $k$ scales. 
We showed that the survey window effect is crucial to take into account to obtain an unbiased estimate of the possible PNG signal. 
A joint cosmology analysis of the three-dimensional power spectra of the galaxy distribution (scalar observables) and the projected tensor observables would also be an interesting direction to explore. 
To do such joint analysis we need both imaging and spectroscopic data for the same cosmological volume, where the imaging data is needed to characterize shapes of individual galaxies and the spectroscopic data is needed to measure distances to galaxies for the 3D power spectrum analysis. 
This is indeed the case for the SDSS BOSS, DESI, Subaru HSC/PFS, VRO LSST/Euclid and Roman Space Telescope. 
We believe that the method developed in this paper helps to extract as much cosmological information as possible from these current and upcoming datasets.

\acknowledgments
We thank Kazu~Akitsu, Zvonimir~Vlah, Yosuke Kobayashi and Sunao Sugiyama for useful discussion, and also thank the Yukawa Institute for Theoretical Physics at Kyoto University for their warm hospitality, where this work was partly done during the YITP-T-21-06 workshop on ``Galaxy shape statistics and cosmology''.
This work was supported in part by World Premier International Research Center Initiative (WPI Initiative), MEXT, Japan, and JSPS KAKENHI Grants No. JP20J22055, No. JP20H05850, No. JP20H05855, No. JP19H00677, and by Basic Research Grant (Super AI) of Institute for AI and Beyond of the University of Tokyo. 
T.K. is supported by JSPS Research Fellowship for Young Scientists. 
The simulations for this work were carried out on Cray XC50 at Center for Computational Astrophysics, National Astronomical Observatory of Japan.

\bibliography{analysis_tensor}

\appendix

\section{\label{sec:relation_full_and_projected_spectrum} Relation between full tensor power spectra and projected tensor power spectra}

We derive the relations between the full tensor power spectra of $s_{ij}(\bk)$ and the projected tensor power spectra of $\gamma(\bk;\hbn) \equiv  e^{(+2)}_{ij}(\hbn) s_{ij}(\bk)$ in Eqs.~(\ref{eq:3D2Drelation_Pcross})--(\ref{eq:3D2Drelation_P-}) where $\hbn$ is the LOS direction.
For the cross spectrum, considering the projection of the definition of the full 3D spectrum (Eq.~\ref{eq:def_of_full_cross_power}), we have
\begin{align}
    \langle \gamma(\bk;\hbn)~ \delta(\bk') \rangle \equiv (2\pi)^3 \delta^3_D(\bk+\bk') e^{(+2)}_{ij}(\hbn) \varepsilon^{(0)}_{ij}(\hbk)~ P_{s\delta}^{(0)}(\bk). 
    \label{eq:app_cross_tmp1}
\end{align}
The above trace of two tensors is given as 
\begin{align}
    e^{(+2)}_{ij}(\hbn) \varepsilon^{(0)}_{ij}(\hbk) \equiv \sqrt{\frac{3}{2}} e^{(+2)}_{ij}(\hbn) \left(\hk_i\hk_j-\frac{1}{3}\delta^K_{ij}\right) = \sqrt{\frac{3}{2}} e^{(+2)}_{ij}(\hbn) \hk_i\hk_j = \sqrt{\frac{3}{8}} (1-\mu^2) e^{2i\phi_{\hbk,\hbn}},
\end{align}
where $\mu \equiv \hbk \cdot \hbn$.
We have used $e^{(+2)}_{ii}(\hbn)=0$ and the definition of the phase factor (Eq.~\ref{eq:def_of_phase_factor}).
Hence, comparing Eq.~(\ref{eq:def_of_prj_cross_power}) with Eq.~(\ref{eq:app_cross_tmp1}), we obtain Eq.~(\ref{eq:3D2Drelation_Pcross}):
\begin{align}
    P_{\gamma \delta}(\bk) &= \sqrt{\frac{3}{8}} (1-\mu_k^2) P_{s\delta}^{(0)}(\bk).
\end{align}

For the auto spectra, $P_\pm (\bk)$ are defined by Eqs.~(\ref{eq:def_of_prj_auto_corr_+}) and (\ref{eq:def_of_prj_auto_corr_-}).
Using these equations and multiplying the full tensor auto spectra (Eq.~\ref{eq:def_of_full_tensor_auto_spectrum}) by $e^{(+2)}_{ij}(\hbn)e^{(\mp2)}_{kl}(\hbn)$, we find
\begin{align}
    \langle \gamma(\bk;\hbn)~ \gamma^*(\bk';\hbn) \rangle \equiv (2\pi)^3 \delta^3_D(\bk+\bk') e^{(+2)}_{ij}(\hbn)e^{(-2)}_{kl}(\hbn) 
    \left\{ \Lambda^{(0)}_{ij,kl}(\hbk)~ P_{ss}^{(0)}(\bk) + \sum_{\lambda=1}^2 \Lambda^{(\lambda)}_{ij,kl}(\hbk)~ \frac{P_{ss}^{(\lambda)}(\bk)}{2} \right\}, \label{eq:app_auto_+_tmp1}\\
    \langle \gamma(\bk;\hbn)~ \gamma(\bk';\hbn) \rangle \equiv (2\pi)^3 \delta^3_D(\bk+\bk') e^{(+2)}_{ij}(\hbn)e^{(+2)}_{kl}(\hbn) 
    \left\{ \Lambda^{(0)}_{ij,kl}(\hbk)~ P_{ss}^{(0)}(\bk) + \sum_{\lambda=1}^2 \Lambda^{(\lambda)}_{ij,kl}(\hbk)~ \frac{P_{ss}^{(\lambda)}(\bk)}{2} \right\}, \label{eq:app_auto_-_tmp1}
\end{align}
thus we need to calculate $e^{(+2)}_{ij}(\hbn)e^{(\mp2)}_{kl}(\hbn) \Lambda^{(\lambda)}_{ij,kl}(\hbk)$ for each $\lambda=0,1,2$. 
For $\lambda=0$ (Eq.~\ref{eq:Lambda0_def}), we can calculate them in a similar way to the cross spectrum as
\begin{align}
    e^{(+2)}_{ij}(\hbn)e^{(-2)}_{kl}(\hbn) \Lambda^{(0)}_{ij,kl}(\hbk) &= \frac{3}{8} e^{(+2)}_{ij}(\hbn)e^{(-2)}_{kl}(\hbn) \hk_i\hk_j\hk_k\hk_l = \frac{3}{8} (1-\mu_k^2)^2, \\
    e^{(+2)}_{ij}(\hbn)e^{(+2)}_{kl}(\hbn) \Lambda^{(0)}_{ij,kl}(\hbk) &= \frac{3}{8} e^{(+2)}_{ij}(\hbn)e^{(+2)}_{kl}(\hbn) \hk_i\hk_j\hk_k\hk_l = \frac{3}{8} (1-\mu_k^2)^2 e^{4i\phi_{\hbk,\hbn}}.
\end{align}
For the ``$+$'' component of $\lambda=1$ (Eq.~\ref{eq:Lambda1_def}), 
\begin{align}
    e^{(+2)}_{ij}(\hbn)e^{(-2)}_{kl}(\hbn) \Lambda^{(1)}_{ij,kl}(\hbk) 
    &= \frac{1}{2} e^{(+2)}_{ij}(\hbn)e^{(-2)}_{kl}(\hbn) \left( \mathcal{P}_{ik}(\hbk)\hk_j\hk_l + \mathcal{P}_{il}(\hbk)\hk_j\hk_k + \mathcal{P}_{jk}(\hbk)\hk_i\hk_l + \mathcal{P}_{jl}(\hbk)\hk_i\hk_k \right) ~\nonumber\\
    &= \frac{1}{2} \mathcal{P}_{ik}(\hbn)\mathcal{P}_{jl}(\hbn) \mathcal{P}_{ik}(\hbk)\hk_j\hk_l ~\nonumber\\
    &= \frac{1}{2} (1-\mu^2) (1+\mu^2) 
    = \frac{1}{4} (1-\mu^2) \left\{(1-\mu)^2 + (1+\mu)^2 \right\}.
\end{align}
We have used the definition of $e^{(\pm 2)}_{ij}(\hbn) \equiv \he^{(\pm 1)}_i(\hbn) \he^{(\pm 1)}_j(\hbn)$ (Eq.~\ref{eq:def_of_eij_+2}) and the projection tensor: 
\begin{align}
    \mathcal{P}_{ij}(\hbn) = \sum_{m=\pm 1} \he^{(m)}_i(\hbn) \he^{(m)*}_j(\hbn),
\end{align}
where $\hbe^{(\pm1)}(\hbn)$ defined in Eq.~(\ref{eq:def_polarization_vector}) is the polarization vector with respect to $\hbn$.
For the ``$-$'' component of $\lambda=1$, 
\begin{align}
    e^{(+2)}_{ij}(\hbn)e^{(+2)}_{kl}(\hbn) \Lambda^{(1)}_{ij,kl}(\hbk) 
    &= \frac{1}{2} e^{(+2)}_{ij}(\hbn)e^{(+2)}_{kl}(\hbn) \left( \mathcal{P}_{ik}(\hbk)\hk_j\hk_l + \mathcal{P}_{il}(\hbk)\hk_j\hk_k + \mathcal{P}_{jk}(\hbk)\hk_i\hk_l + \mathcal{P}_{jl}(\hbk)\hk_i\hk_k \right) ~\nonumber\\
    &= 2e^{(+2)}_{ij}(\hbn)e^{(+2)}_{kl}(\hbn) \mathcal{P}_{ik}(\hbk)\hk_j\hk_l ~\nonumber\\
    &= -2e^{(+2)}_{ij}(\hbn)e^{(+2)}_{kl}(\hbn) \hk_i\hk_k\hk_j\hk_l ~\nonumber\\
    &= - \frac{1}{2} (1-\mu^2)^2 e^{4i\phi_{\hbk,\hbn}}.
\end{align}
For $\lambda=2$ (Eq.~\ref{eq:Lambda2_def}), 
\begin{align}
    e^{(+2)}_{ij}(\hbn)e^{(-2)}_{kl}(\hbn) \Lambda^{(2)}_{ij,kl}(\hbk) 
    &= \frac{1}{2} e^{(+2)}_{ij}(\hbn)e^{(-2)}_{kl}(\hbn) \left(  \mathcal{P}_{ik}(\hbk)\mathcal{P}_{jl}(\hbk) + \mathcal{P}_{il}(\hbk)\mathcal{P}_{jk}(\hbk) - \mathcal{P}_{ij}(\hbk)\mathcal{P}_{kl}(\hbk) \right) ~\nonumber\\
    &= \frac{1}{4} \mathcal{P}_{ik}(\hbn)\mathcal{P}_{jl}(\hbn) \mathcal{P}_{ik}(\hbk) \mathcal{P}_{jl}(\hbk) - \frac{1}{2}e^{(+2)}_{ij}(\hbn)e^{(-2)}_{kl}(\hbn) \hk_i\hk_j\hk_k\hk_l ~\nonumber\\
    &= \frac{1}{4} (1+\mu^2)^2 - \frac{1}{8} (1-\mu^2)^2 
    = \frac{1}{16} \left\{(1-\mu)^4 + (1+\mu)^4 \right\}. ~\\
    e^{(+2)}_{ij}(\hbn)e^{(+2)}_{kl}(\hbn) \Lambda^{(2)}_{ij,kl}(\hbk) 
    &= \frac{1}{2} e^{(+2)}_{ij}(\hbn)e^{(+2)}_{kl}(\hbn) \left(  \mathcal{P}_{ik}(\hbk)\mathcal{P}_{jl}(\hbk) + \mathcal{P}_{il}(\hbk)\mathcal{P}_{jk}(\hbk) - \mathcal{P}_{ij}(\hbk)\mathcal{P}_{kl}(\hbk) \right) ~\nonumber\\
    &= \frac{1}{2} e^{(+2)}_{ij}(\hbn)e^{(+2)}_{kl}(\hbn) \hk_i\hk_j\hk_k\hk_l ~\nonumber\\
    &= \frac{1}{8} (1-\mu^2)^2  e^{4i\phi_{\hbk,\hbn}}.
\end{align}
Substituting all results into Eqs.~(\ref{eq:app_auto_+_tmp1}) and (\ref{eq:app_auto_-_tmp1}) and comparing them with Eqs.~(\ref{eq:def_of_prj_auto_power_+}) and (\ref{eq:def_of_prj_auto_power_-}), respectively, we obtain the relations (Eqs.~\ref{eq:3D2Drelation_P+} and \ref{eq:3D2Drelation_P-}):
\begin{align}
    P_{+}(\bk) &= \frac{3}{8} (1-\mu_k^2)^2 P_{ss}^{(0)}(\bk) + \frac{1}{8} (1-\mu_k^2)\{ (1-\mu_k)^2 + (1+\mu_k)^2\} P_{ss}^{(1)}(\bk) + \frac{1}{32} \{ (1-\mu_k)^4 + (1+\mu_k)^4 \} P_{ss}^{(2)}(\bk), \\
    P_{-}(\bk) &= (1-\mu_k^2)^2 \left[ \frac{3}{8} P_{ss}^{(0)}(\bk) - \frac{1}{4} P_{ss}^{(1)}(\bk) + \frac{1}{16} P_{ss}^{(2)}(\bk) \right]. 
\end{align}

\section{\label{sec:calc_auto_est}Estimator of auto-power spectra}
We derive the FFT-based estimators for the auto-power spectra defined in Eqs.~(\ref{eq:lpp_estimator_auto_+}) and (\ref{eq:lpp_estimator_auto_-}). 
In the case of the ``plus'' component, $P_+(\bk)$ (Eq.~\ref{eq:relation_auto_gpp_+}) 
we define the multipole moments with respect to the usual Legendre polynomials as in the case of the clustering:
\begin{align}
    \rvred{\hat{P}}^{(\ell)}_+(k)
    &\equiv (2\ell+1) \int_{\hbk,\bx,\bx'} \gamma(\bx) \gamma^*(\bx') e^{-i\bk \cdot (\bx - \bx')} \mathcal{L}_\ell (\hbk \cdot \hbd) ~\nonumber\\
    &\simeq (2\ell+1) \int_{\hbk} \left[ \int_{\hbx} \gamma(\bx) e^{-i\bk \cdot \bx}  \mathcal{L}_{\ell} (\hbk \cdot \hbx)  \right] 
    \left[ \int_{\hbx'} \gamma^*(\bx') e^{i\bk \cdot \bx'} \right] ~\nonumber\\
    &\equiv (2\ell+1) \int_{\hbk} \gamma^{(\ell)}(\bk)~ \gamma^{*}(-\bk). 
    \label{eq:app_lpp_est_auto_plus}
\end{align}
On the other hand, the ``minus'' component,  $P_-(\bk)$, considering the phase factor in Eq.~(\ref{eq:relation_auto_gpp_-}), we define the multipole moments with respect to the associated Legendre expansion of order $m=4$:
\begin{align}
    \rvred{\hat{P}}^{(L)}_-(k)
    &\equiv (2L+1) \frac{(L-4)!}{(L+4)!} \int_{\hbk,\bx,\bx'} \gamma(\bx) \gamma(\bx')  e^{-4i\phi_{\hbk,\hbd}} e^{-i\bk \cdot (\bx - \bx')} \mathcal{L}^{m=4}_L (\hbk \cdot \hbd) ~\nonumber\\
    &= (2L+1) \frac{(L-4)!}{(L+4)!} \int_{\hbk,\bx,\bx'} \gamma(\bx) \gamma(\bx') 4e^*_{ij}(\hbd)e^*_{kl}(\hbd)\hk_i\hk_j\hk_k\hk_l e^{-i\bk \cdot (\bx - \bx')} \frac{\mathcal{L}^{m=4}_{L} (\hbk \cdot \hbd)}{[1-(\hbk\cdot\hbd)^2]^2} ~\nonumber\\
    &\simeq (2L+1) \frac{(L-4)!}{(L+4)!} \int_{\hbk} \left[ \int_{\hbx} \gamma(\bx) 4e^*_{ij}(\hbx) e^*_{kl}(\hbx) e^{-i\bk \cdot \bx} \frac{\mathcal{L}^{m=4}_{L} (\hbk \cdot \hbx)}{[1-(\hbk\cdot\hbx)^2]^2} \right] \hk_i\hk_j\hk_k\hk_l
    \left[ \int_{\hbx'} \gamma(\bx') e^{i\bk \cdot \bx'} \right] ~\nonumber\\
    &\equiv (2L+1) \frac{(L-4)!}{(L+4)!} \int_{\hbk} \Xi^{(\ell)}_{ijkl}(\bk) \hk_i\hk_j \hk_k\hk_l ~\gamma(-\bk).
    \label{eq:app_lpp_est_auto_minus}
\end{align}
Note that we have used the endpoint approximation; $\hbd \simeq \hbx$, in the second line of Eq.~(\ref{eq:app_lpp_est_auto_plus}) and the third line of Eq.~(\ref{eq:app_lpp_est_auto_minus}).

\section{\label{sec:hankel_transform}Derivation of Hankel transforms}
First, we define the local cross spectrum as the Fourier transform of the cross correlation function:
\begin{align}
    P_{\gamma \delta}(\bk, \hbd) \equiv \int_{\br} \xi_{\gamma \delta} (r, \hbr \cdot \hbd) e^{2i\phi_{\hbr,\hbd} - 2i\phi_{\hbk,\hbd}} e^{-i\bk \cdot \br}.
    \label{eq:def_of_cross_power}
\end{align}
Assuming the cross correlation function can be expanded in terms of the associated Legendre polynomials with $m=2$:
\begin{align}
    \xi_{\gamma \delta} (r, \hbr \cdot \hbd) \equiv \sum^{\infty}_{L=2} \xi^{(L)}_{\gamma \delta}(r) \mathcal{L}^{m=2}_{L} (\hbr \cdot \hbd),
    \label{eq:xi_multipole_expansion}
\end{align}
and using the plane-wave expansion and the addition theorem for the Legendre polynomials:
\begin{align}
    e^{i\bk \cdot \br} &= \sum^{\infty}_{q=0} (2q+1) i^q j_q (kr) \mathcal{L}_{q} (\hbk \cdot \hbr), \label{eq:plane-wave_expansion}\\
    \mathcal{L}_{q} (\hbk \cdot \hbr) &= \frac{4\pi}{2q+1} \sum_{n=-q}^{q} Y_q^n (\hbk) Y_q^{n*} (\hbr), 
    \label{eq:legendre_harmonics_expansion}
\end{align}
with the rotational invariance of the inner product, $\hbk \cdot \hbr = (R\hbk) \cdot (R\hbr)$, for any rotation matrix $R$, Eq.~{(\ref{eq:def_of_cross_power})} becomes
\begin{align}
    P_{\gamma \delta}(\bk, \hbd) &= \sum^{\infty}_{L=2} \sum^{\infty}_{q=0} \sum_{n=-q}^{q} (-i)^q Y_q^n (S^{-1}(\hbd)\hbk) e^{-2i\phi_{\hbk,\hbd}}
    4\pi \int \mathrm{d}\Omega_{\hbr}~ \mathcal{L}^{m=2}_{L}(\hbr \cdot \hbd) e^{2i\phi_{\hbr,\hbd}} Y_q^{n*} (S^{-1}(\hbd)\hbr)
    \int r^2 \mathrm{d}r~ \xi^{(L)}_{\gamma \delta}(r) j_q(kr) \nonumber \\
    &= \sum^{\infty}_{L=2} \sum^{\infty}_{q=0} \sum_{n=-q}^{q} (-i)^q Y_q^n (S^{-1}(\hbd)\hbk) e^{-2i\phi_{\hbk,\hbd}} 
    4\pi \int \mathrm{d}\Omega_{\hbr}~ \frac{Y_L^{m=2} (S^{-1}(\hbd)\hbr)}{\mathcal{N}^{m=2}_{L}} Y_q^{n*} (S^{-1}(\hbd)\hbr)
    \int r^2 \mathrm{d}r~ \xi^{(L)}_{\gamma \delta}(r) j_q(kr) \nonumber \\
    &= \sum^{\infty}_{L=2} \sum^{\infty}_{q=0} \sum_{n=-q}^{q} (-i)^q Y_q^n (S^{-1}(\hbd)\hbk) e^{-2i\phi_{\hbk,\hbd}} 
    \frac{4\pi \delta_{L q} \delta_{n2}}{\mathcal{N}^{m=2}_{L}}
    \int r^2 \mathrm{d}r~ \xi^{(L)}_{\gamma \delta}(r) j_q(kr) \nonumber \\
    &= \sum^{\infty}_{L=2} \mathcal{L}^{m=2}_{L}(\hbk \cdot \hbd)  4\pi (-i)^L \int r^2 \mathrm{d}r~ \xi^{(L)}_{\gamma \delta}(r) j_L(kr),
    \label{eq:power_xi_relation_app}
\end{align}
where $S(\hbd)$ is the standard rotation matrix that takes $\hbx_3$ (3-axis) into the arbitrary direction $\hbd$:
\begin{equation}
    S_{ij}(\hbd) \equiv
    \begin{pmatrix}
        \cos{\theta}\cos{\phi} & -\sin{\phi} & \sin{\theta}\cos{\phi}\\
        \cos{\theta}\sin{\phi} & \cos{\phi} & \sin{\theta}\sin{\phi}\\
        -\sin{\theta} & 0 & \cos{\theta}
    \end{pmatrix},
    \label{eq:def_standard_rotation}
\end{equation}
for $\hbd \equiv (\sin{\theta}\cos{\phi}, \sin{\theta}\sin{\phi}, \cos{\theta})$, and $\mathcal{N}^{m}_{\ell}$ is the normalization factor in the definition of the spherical harmonics:
\begin{align}
    Y_\ell^m(\hba) = \mathcal{N}^{m}_{\ell} \mathcal{L}^{m}_{\ell}(\hba \cdot \hbx_3) e^{im\phi_{\hba,\hbx_3}}.
    \label{eq:def_spherical_harmonics}
\end{align}
We have also used the orthogonality of the spherical harmonics:
\begin{align}
    \int_{\hba} Y_\ell^{m} (\hba) Y_q^{n*} (\hba) = \frac{1}{4\pi} \delta_{\ell q} \delta_{mn},
    \label{eq:orthogonality_harmonics}
\end{align}
and the following identity outlined in Appendix~\ref{sec:derivations}:
\begin{align}
    Y_\ell^m(S^{-1}(\hbb)\hba) = \mathcal{N}^{m}_{\ell} \mathcal{L}^{m}_{\ell}(\hba \cdot \hbb) e^{im\phi_{\hba,\hbb}}.
    \label{eq:spherical_harmonics_identity}
\end{align}
From Eq.~(\ref{eq:power_xi_relation_app}), we can define the multipole moments of the cross power spectrum as
\begin{align}
    P^{(L)}_{\gamma \delta}(k) &\equiv (2L+1) \frac{(L-2)!}{(L+2)!} \int_{\hbk} P_{\gamma \delta}(\bk, \hbd) \mathcal{L}^{m=2}_{L}(\hbk \cdot \hbd) \nonumber\\
    &= 4\pi (-i)^L \int r^2 \mathrm{d}r~ \xi^{(L)}_{\gamma \delta}(r) j_L(kr), 
    \label{eq:pmulti_ximulti_relation}
\end{align}
where we have used the orthogonality of the associated Legendre polynomials (Eq.~\ref{eq:ortho_asso_legendre}):
\begin{align}
    \int_{-1}^1 \frac{\mathrm{d}\mu}{2}~ \mathcal{L}^{m}_{L}(\mu) \mathcal{L}^{m}_{L'}(\mu) = \frac{(L+m)!}{(2L+1)(L-m)!} \delta_{L L'}.
    \label{eq:asso_legendre_ortho}
\end{align}

\section{Unified Formula of Window Convolutions}
\label{sec:window_convolution}
We first review the window convolution formula of the galaxy clustering shown in Refs.~\citep{Wilson+2017,Beutler+2017,Beutler&McDonald2021:wide_angle} and next slightly generalize it to derive that of the projected tensor power spectrum.

\subsection{Galaxy clustering}
\label{subsec:review_of_clustering_convolution}
The window-convolved power spectrum of galaxy clustering may be written by
\begin{align}
    \tilde{P}(\bk,\hbd) \equiv \int_{\br} \xi(\br,\hbd) Q(\br,\hbd) e^{-i\bk \cdot \br} = \int_{\bk'} |W(\bk-\bk',\hbd)|^2 P(\bk',\hbd).
\end{align}
The multipole moments of it is thus defined as
\begin{align}
    \tilde{P}^{(\ell)}(k) 
    &\equiv (2\ell+1) \int_{\hbk} \tilde{P}(\bk,\hbd) \mathcal{L}_\ell(\hbk\cdot\hbd) \nonumber\\
    &= (2\ell+1) \int_{\hbk, \hbk'} |W(\bk-\bk',\hbd)|^2 P(\bk',\hbd) \mathcal{L}_\ell(\hbk\cdot\hbd). \label{eq:p_dd_ell_1}
\end{align}
Expanding the underlying power spectrum and the window auto-correlation function in terms of the Legendre polynomials:
\begin{align}
    P(\bk',\hbd) &\equiv \sum_{\ell'} P^{(\ell')}(k') {\cal L}_{\ell'}(\hbk'\cdot\hbd), \\
    |W(\bk-\bk',\hbd)|^2 &= \int_{\br} Q(\br,\hbd) e^{i(\bk-\bk')\cdot\br} \equiv \int_{\br} \sum_{\ell''} Q_{\ell''}(r) \mathcal{L}_{\ell''}(\hbr \cdot \hbd) e^{i(\bk-\bk')\cdot\br}, \label{eq:app_W2_multipole_expansion}
\end{align}
and also using the plane-wave expansion (Eq.~\ref{eq:plane-wave_expansion}):
\begin{align}
    e^{i(\bk-\bk')\cdot\br} = \sum_{p,q} (2p+1)(2q+1)(-i)^{p}i^q j_p(kr)j_q(k'r) \mathcal{L}_{p}(\hbk \cdot \hbr) \mathcal{L}_{q}(\hbk' \cdot \hbr),
\end{align}
we can rewrite Eq.~(\ref{eq:p_dd_ell_1}) as
\begin{align}
    \tilde{P}^{(\ell)}(k) &= (2\ell+1) \sum_{\ell',\ell'',p,q} \int_{\hbk, \hbk', \br} P^{(\ell')}(k') Q_{\ell''}(r) (2p+1)(2q+1)(-i)^{p}i^q j_p(kr)j_q(k'r) ~\nonumber\\
    &\qquad \qquad \qquad \qquad \qquad \times \mathcal{L}_\ell(\hbk\cdot\hbd) {\cal L}_{\ell'}(\hbk'\cdot\hbd) \mathcal{L}_{\ell''}(\hbr \cdot \hbd) \mathcal{L}_{p}(\hbk \cdot \hbr) \mathcal{L}_{q}(\hbk' \cdot \hbr).
\end{align}
Using the angle integral of the product of two Legendre polynomials:
\begin{align}
    \int_{\hbk} \mathcal{L}_{\ell_1}(\hbk \cdot \hba) \mathcal{L}_{\ell_2}(\hbk \cdot \hbb) = \frac{\delta_{\ell_1 \ell_2}}{2\ell_1+1} \mathcal{L}_{\ell_1}(\hba \cdot \hbb),
    \label{eq:legendre_identity}
\end{align}
we do the $\hbk$ and $\hbk'$ integrals and then obtain
\begin{align}
    \tilde{P}^{(\ell)}(k) = (2\ell+1) \sum_{\ell',\ell''} \int \frac{k'^2\mathrm{d}k'}{2\pi^2} \int_{\br} P^{(\ell')}(k') Q_{\ell''}(r) (-i)^{\ell}i^{\ell'} j_\ell(kr)j_{\ell'}(k'r) \mathcal{L}_\ell(\hbr\cdot\hbd) {\cal L}_{\ell'}(\hbr\cdot\hbd) \mathcal{L}_{\ell''}(\hbr \cdot \hbd).
\end{align}
Finally by using the Gaunt integral:
\begin{align}
    \int\mathrm{d}\Omega_{\hbr} 
    Y_{\ell_1}^{m_1}(\hbr) Y_{\ell_2}^{m_2}(\hbr) Y_{\ell_3}^{m_3}(\hbr) = \sqrt{\frac{(2\ell_1+1)(2\ell_2+1)(2\ell_3+1)}{4\pi}} 
    \begin{pmatrix}
        \ell_1 & \ell_2 & \ell_3\\
        0 & 0 & 0
    \end{pmatrix}
    \begin{pmatrix}
        \ell_1 & \ell_2 & \ell_3\\
        m_1 & m_2 & m_3
    \end{pmatrix}, 
    \label{eq:gaunt_integral}
\end{align}
where 
$\begin{pmatrix}
        \ell_1 & \ell_2 & \ell_3\\
        m_1 & m_2 & m_3
\end{pmatrix}$
is the Wigner $3j$ symbol, we obtain the fully angle-integrated result:
\begin{align}
    \tilde{P}^{(\ell)}(k) &= (2\ell+1) \sum_{\ell',\ell''} \int \frac{2}{\pi} k'^2\mathrm{d}k' \int r^2\mathrm{d}r P^{(\ell')}(k') Q_{\ell''}(r) (-i)^{\ell}i^{\ell'} j_\ell(kr)j_{\ell'}(k'r) 
    \begin{pmatrix}
        \ell'' & \ell & \ell'\\
        0 & 0 & 0
    \end{pmatrix}^2 ~\nonumber\\
    &= 4\pi (-i)^{\ell} \int r^2\mathrm{d}r j_\ell(kr) \sum_{\ell'} 
    \left[\sum_{\ell''} Q_{\ell''}(r) (2\ell+1) 
    \begin{pmatrix}
        \ell'' & \ell & \ell'\\
        0 & 0 & 0
    \end{pmatrix}^2 \right]
    \left[i^{\ell'} \int \frac{k'^2\mathrm{d}k'}{2\pi^2} j_{\ell'}(k'r) P^{(\ell')}(k') \right].
    \label{eq:app_convolution_clustering_case}
\end{align}

\subsection{Generalization to the projected tensor power spectrum}
\label{subsec:derivation_of_general_convolution}

If we have the projected tensor field, $\gamma(\bx)$, we can think the cross-, plus- and minus-power spectra, $(P_{\gamma\delta}, P_+, P_-)$ in addition to the galaxy clustering power spectrum $P(=P_{\delta\delta})$. 
In the case of $P_+$, the window convolution is the same as the clustering case:
\begin{align}
    \tilde{P}_+(\bk,\hbd) \equiv \int_{\br} \xi_+(\br,\hbd) Q(\br,\hbd) e^{-i\bk \cdot \br} = \int_{\bk'} |W(\bk-\bk',\hbd)|^2 P_+(\bk',\hbd),
\end{align}
hence the resulting convolution expression is exactly the same as Eq.~(\ref{eq:app_convolution_clustering_case}).

To consider the case of the cross spectrum and the ``minus'' component of the auto spectrum, we slightly generalize the previous result beginning with the fact that the window effect on any correlation function is defined by $\tilde{\xi}_{\rm X}(\br) \equiv \xi_{\rm X}(\br) Q(\br)$ where ${\rm X}$ is the label of the statistics.
The definitions of the convolution thus should be
\begin{align}
    \tilde{P}_{\rm X}(\bk,\hbd) \equiv \int_{\br} \xi_{\rm X}(\br,\hbd) Q(\br,\hbd) e^{im_{\rm X}\phi_{\hbr,\hbd} - im_{\rm X}\phi_{\hbk,\hbd}}  e^{-i\bk \cdot \br} = \int_{\bk'} |W(\bk-\bk',\hbd)|^2 e^{-im_{\rm X}\phi_{\hbk,\hbd} + im_{\rm X}\phi_{\hbk',\hbd}} P_{\rm X}(\bk',\hbd),
\end{align}
where $e^{im_{\rm X}\phi}$ is the respective phase factor; $({\rm X},m_{\rm X}) \in \left\{(\delta\delta,0), (\gamma\delta, 2), (+,0), (-,4) \right\}$ and the multipole components are defined with respect to the associated Legendre polynomials:
\begin{align}
    \tilde{P}^{(L)}_{\rm X}(k) 
    &\equiv (2L+1) \frac{(L-m_{\rm X})!}{(L+m_{\rm X})!} \int_{\hbk} \tilde{P}_{\rm X}(\bk,\hbd) \mathcal{L}^{m_{\rm X}}_L(\hbk\cdot\hbd) \nonumber\\
    &= (2L+1) \frac{(L-m_{\rm X})!}{(L+m_{\rm X})!} \int_{\hbk, \hbk'} |W(\bk-\bk',\hbd)|^2 e^{-im_{\rm X}\phi_{\hbk,\hbd} + im_{\rm X}\phi_{\hbk',\hbd}} P_{\rm X}(\bk',\hbd) \mathcal{L}^{m_{\rm X}}_L(\hbk\cdot\hbd). \label{eq:p_xx_ell_1}
\end{align}

From this definition, by using the associated Legendre expansion for the underlying theoretical model:
\begin{align}
    P_{\rm X}(\bk',\hbd) \equiv \sum_{L'} P_{\rm X}^{(L')}(k') {\cal L}^{m_{\rm X}}_{L'}(\hbk'\cdot\hbd),
\end{align}
and substituting Eqs.~(\ref{eq:app_W2_multipole_expansion}) and Eq.~(\ref{eq:plane-wave_expansion}), we obtain
\begin{align}
    \tilde{P}^{(L)}_{\rm X}(k) &= (2L+1) \frac{(L-m_{\rm X})!}{(L+m_{\rm X})!} \sum_{L',\ell'',p,q} \int_{\hbk, \hbk', \br} P^{(L')}_{\rm X}(k') Q_{\ell''}(r) (2p+1)(2q+1)(-i)^{p}i^q j_p(kr)j_q(k'r) ~\nonumber\\
    &\qquad \qquad \qquad \qquad \qquad \qquad \qquad \times e^{-im_{\rm X}\phi_{\hbk,\hbd} + im_{\rm X}\phi_{\hbk',\hbd}} \mathcal{L}^{m_{\rm X}}_L(\hbk\cdot\hbd) {\cal L}^{m_{\rm X}}_{L'}(\hbk'\cdot\hbd) \mathcal{L}_{\ell''}(\hbr \cdot \hbd) \mathcal{L}_{p}(\hbk \cdot \hbr) \mathcal{L}_{q}(\hbk' \cdot \hbr).
\end{align}
By using Eqs.~(\ref{eq:def_spherical_harmonics}), (\ref{eq:orthogonality_harmonics}) and (\ref{eq:spherical_harmonics_identity}), we do the 
$\hbk$ and $\hbk'$ integrals:
\begin{align}
    \tilde{P}^{(L)}_{\rm X}(k) = (2L+1) \frac{(L-m_{\rm X})!}{(L+m_{\rm X})!} \sum_{L',\ell''} \int \frac{k'^2\mathrm{d}k'}{2\pi^2} \int_{\br} P^{(L')}_{\rm X}(k') Q_{\ell''}(r) (-i)^{L}i^{L'} j_L(kr)j_{L'}(k'r) \mathcal{L}^{m_{\rm X}}_\ell(\hbr\cdot\hbd) {\cal L}^{m_{\rm X}}_{L'}(\hbr\cdot\hbd) \mathcal{L}_{\ell''}(\hbr \cdot \hbd),
\end{align}
and also by using Eq.~(\ref{eq:gaunt_integral}), we finally obtain
\begin{align}
    \tilde{P}^{(L)}_{\rm X}(k) &= (2L+1) \frac{(L-m_{\rm X})!}{(L+m_{\rm X})!} \sum_{L',\ell''} \int \frac{2}{\pi} k'^2\mathrm{d}k' \int r^2\mathrm{d}r P^{(L')}_{\rm X}(k') Q_{\ell''}(r) (-i)^{L}i^{L'} j_L(kr)j_{L'}(k'r) ~\nonumber\\
    &\qquad \qquad \qquad \qquad \qquad \qquad \qquad \times \sqrt{\frac{(L+m_{\rm X})!}{(L-m_{\rm X})!} \frac{(L'+m_{\rm X})!}{(L'-m_{\rm X})!}}
    \begin{pmatrix}
        \ell'' & L & L'\\
        0 & 0 & 0
    \end{pmatrix}
    \begin{pmatrix}
        \ell'' & L & L'\\
        0 & m_{\rm X} & -m_{\rm X}
    \end{pmatrix} ~\nonumber\\
    &= 4\pi (-i)^{L} \int r^2\mathrm{d}r j_L(kr) \sum_{L'} 
    \left[\sum_{\ell''} Q_{\ell''}(r) (2L+1) \sqrt{\frac{(L-m_{\rm X})!}{(L+m_{\rm X})!} \frac{(L'+m_{\rm X})!}{(L'-m_{\rm X})!}}
    \begin{pmatrix}
        \ell'' & L & L'\\
        0 & 0 & 0
    \end{pmatrix}
    \begin{pmatrix}
        \ell'' & L & L'\\
        0 & m_{\rm X} & -m_{\rm X}
    \end{pmatrix} \right] ~\nonumber\\
    &\qquad \qquad \qquad \qquad \qquad \qquad \qquad \times \left[i^{L'} \int \frac{k'^2\mathrm{d}k'}{2\pi^2} j_{L'}(k'r) P^{(L')}_{\rm X}(k') \right].
    \label{eq:app_convolution_unified}
\end{align}
This corresponds to Eqs.~(\ref{eq:window_convolution_X}) and (\ref{eq:def_of_Q_llp}) in the main text.

\section{\label{sec:derivations}Derivations of some formulae}
\subsection{\label{subsec:derivation_eqC}Derivation of Eq.~(\ref{eq:spherical_harmonics_identity})}
Comparing the goal, Eq.~(\ref{eq:spherical_harmonics_identity}), with the definition of the spherical harmonics, Eq.~(\ref{eq:def_spherical_harmonics}), we first show the two identities:
\begin{align*}
    \begin{cases}
        \hbu(\hba, \hbb) \cdot \hbx_3 &= \hba \cdot \hbb, \\
        e^{im\phi_{\hbu(\hba, \hbb), \hbx_3}} &= e^{im\phi_{\hba,\hbb}},
    \end{cases}
\end{align*}
where $\hbu(\hba, \hbb) \equiv S^{-1}(\hbb)\hba$.
For the first equation, from the definition of the standard rotation matrix $S$ in Eq.~(\ref{eq:def_standard_rotation}), we have $\hbu(\hba, \hbb) \cdot \hbx_3 = \left( S^{-1}(\hbb)\hba \right) \cdot \hbx_3 = \hba \cdot \left( S(\hbb)\hbx_3 \right) = \hba \cdot \hbb$.
We write down the phase factor for arbitrary $m$ as
\begin{align}
    e^{im\phi_{\hbu,\hbx_3}} = \left(e^{i\phi_{\hbu,\hbx_3}}\right)^m 
    = \left( \frac{\sqrt{2} \he^{(+1)}_i(\hbx_3)\hu_i}{\left[ \mathcal{P}_{ij}(\hbx_3)\hu_i\hu_j \right]^{1/2}} \right)^m,
\end{align}
where $\hbe^{(+1)}(\hbx_3)$ is the polarization vector defined in Eq.~(\ref{eq:def_polarization_vector}).
For the numerator, 
\begin{align*}
    \hbe^{(+1)}(\hbx_3) \cdot \hbu(\hba, \hbb) = \hbe^{(+1)}(\hbx_3) \cdot \left( S^{-1}(\hbb)\hba \right) = \left( S(\hbb)\hbe^{(+1)}(\hbx_3) \right) \cdot \hba = \hbe^{(+1)}(\hbb) \cdot \hba,
\end{align*}
in the same way as the first equation.
Also for the denominator, we obtain
\begin{align*}
    \mathcal{P}_{ij}(\hbx_3)\hu_i\hu_j 
    &\equiv \left( \delta_{ij} - \hz_i \hz_j \right) S^{-1}_{ii'}(\hbb)\ha_{i'} S^{-1}_{jj'}(\hbb)\ha_{j'} \\
    &= \left( S^{-1}_{ii'}(\hbb)S^{-1}_{ij'}(\hbb) - S^{-1}_{ii'}(\hbb)\hz_i S^{-1}_{jj'}(\hbb)\hz_j \right) \ha_{i'}\ha_{j'} \\
    &= \left( S_{i'i}(\hbb)S^{-1}_{ij'}(\hbb) - S_{i'i}(\hbb)\hz_i S_{j'j}(\hbb)\hz_j \right) \ha_{i'}\ha_{j'} \\
    &= \left( \delta_{i'j'} - \hb_{i'} \hb_{j'} \right)\ha_{i'}\ha_{j'} \\
    &= \mathcal{P}_{ij}(\hbb)\ha_i\ha_j. 
\end{align*}
From these, we get
\begin{align}
    e^{im\phi_{\hbu,\hbx_3}}
    = \left( \frac{\sqrt{2} \he^{(+1)}_i(\hbx_3)\hu_i}{\left[ \mathcal{P}_{ij}(\hbx_3)\hu_i\hu_j \right]^{1/2}} \right)^m 
    = \left( \frac{\sqrt{2} \he^{(+1)}_i(\hbb)\ha_i}{\left[ \mathcal{P}_{ij}(\hbb)\ha_i\ha_j \right]^{1/2}} \right)^m 
    = e^{im\phi_{\hba,\hbb}}.
\end{align}
Therefore,
\begin{align*}
    Y_\ell^m(\hbu(\hba, \hbb)) 
    &= \mathcal{N}^{m}_{\ell} \mathcal{L}^{m}_{\ell}(\hbu(\hba, \hbb) \cdot \hbx_3) e^{im\phi_{\hbu(\hba, \hbb),\hbx_3}} \\
    &= \mathcal{N}^{m}_{\ell} \mathcal{L}^{m}_{\ell}(\hba \cdot \hbb) e^{im\phi_{\hba,\hbb}}.
\end{align*}

\subsection{\label{subsec:derivation_eqD}Derivation of Eq.~(\ref{eq:formulae_associatedLegendre})}
\begin{align*}
    \int_{\hbk} e^{im\phi_{\hbk,\hba}} \mathcal{L}^{m}_{\ell}(\hbk \cdot \hba) \mathcal{L}_{\ell'}(\hbk \cdot \hbb) 
    &= \int_{\hbk} \frac{Y_\ell^{m}(S^{-1}(\hba)\hbk)}{\mathcal{N}^{m}_{\ell}} \cdot \frac{4\pi}{2\ell'+1} \sum_{n=-\ell'}^{\ell'} Y_{\ell'}^{n*} (S^{-1}(\hba)\hbk) Y_{\ell'}^{n} (S^{-1}(\hba)\hbb) \\ 
    &= \frac{4\pi}{2\ell'+1} \sum_{n=-\ell'}^{\ell'} \frac{1}{\mathcal{N}^{m}_{\ell}} Y_{\ell'}^{n} (S^{-1}(\hba)\hbb) \int_{\hbk} Y_\ell^{m}(S^{-1}(\hba)\hbk) Y_{\ell'}^{n*} (S^{-1}(\hba)\hbk) \\
    &= \frac{4\pi}{2\ell'+1} \sum_{n=-\ell'}^{\ell'} \frac{1}{\mathcal{N}^{m}_{\ell}} Y_{\ell'}^{n} (S^{-1}(\hba)\hbb) \frac{\delta_{\ell \ell'} \delta_{mn}}{4\pi} \\
    &= \frac{\delta_{\ell \ell'}}{2\ell+1} e^{im\phi_{\hba,\hbb}} \mathcal{L}^{m}_{\ell}(\hba \cdot \hbb).
\end{align*}
We have used Eq.~(\ref{eq:spherical_harmonics_identity}) and Eq.~(\ref{eq:legendre_harmonics_expansion}) in the first line,  Eq.~(\ref{eq:orthogonality_harmonics}) in the third line, and Eq.~(\ref{eq:spherical_harmonics_identity}) again in the last line, respectively.
Taking the complex conjugate of both sides, we obtain Eq.~(\ref{eq:formulae_associatedLegendre}).

\end{document}